\documentclass[preprintnumbers, floatfix, letterpaper, onecolumn,aps,prd,epsfig,nofootinbib,natbib,
longbibliography
]{revtex4-2}
\usepackage{bm,graphicx,dcolumn,epstopdf,epsf, latexsym,mathbbol, amssymb,amsmath,color,slashed, mathrsfs,mathcomp,simplewick}
\usepackage[center]{subfigure}
\usepackage{multirow}
\usepackage{makecell}
\usepackage{epstopdf}
\usepackage[colorlinks,linkcolor=blue,citecolor=blue,urlcolor=blue]{hyperref}
\usepackage{amsmath}
\begin{document}
\allowdisplaybreaks
%%%%%%%%%%%%%%%%%%%%%%%%
 \newcommand{\bq}{\begin{equation}}
 \newcommand{\eq}{\end{equation}}
 \newcommand{\bqn}{\begin{eqnarray}}
 \newcommand{\eqn}{\end{eqnarray}}
 \newcommand{\nb}{\nonumber}
 \newcommand{\lb}{\label}
\newcommand{\f}{\frac}
\newcommand{\p}{\partial}
%%%%%%%%%%%%%%%%%%%%%%%%%
\newcommand{\PRL}{Phys. Rev. Lett.}
\newcommand{\PLB}{Phys. Lett. B}
\newcommand{\PRD}{Phys. Rev. D}
\newcommand{\CQG}{Class. Quantum Grav.}
\newcommand{\JCAP}{J. Cosmol. Astropart. Phys.}
\newcommand{\JHEP}{J. High. Energy. Phys.}
\newcommand{\orcid}[1]{\href{https://orcid.org/#1}{\includegraphics[width=10pt]{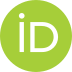}}}
 %%%%%%%%%%%%%%%%%%%%%%%%
\title{The Dynamics of Reheating in Loop Quantum Cosmology}

\author{Yogesh\orcid{0000-0002-7638-3082}$^{1}$}
\email{yogesh@zjut.edu.cn,yogeshjjmi@gmail.com}

\author{Bao-Fei Li$^{1}$}
\email{libaofei@zjut.edu.cn}

\author{Mayukh R. Gangopadhyay\orcid{0000-0002-1466-8525}$^{2}$}
\email{mayukh$_$ccsp@sgtuniversity.org, mayukhraj@gmail.com}

\author{Anzhong Wang\orcid{0000-0002-8852-9966}$^{3}$}
\email{Anzhong$_$Wang@baylor.edu; the corresponding author}

\affiliation{$^{1}$ Institute for Theoretical Physics and Cosmology, Zhejiang University of Technology, Hangzhou, 310032, China\\
$^{2}$ Centre for Cosmology and Science Popularization, SGT University, Gurugram, Haryana-122505, India\\ 
$^{3}$ GCAP-CASPER, Department of Physics $\&$ Astronomy, Baylor University, Waco, TX 76798-7316, USA}
\date{\today}

\begin{abstract}

In loop quantum cosmology (LQC), the initial singularity is replaced by a quantum bounce, leading to a universal post-bounce evolution characterized by three distinct epochs: bouncing, transition, and slow-roll inflation, before the hot big-bang universe starts. While the generic nature of inflation in LQC is well-established, the subsequent reheating phase—the process that thermalizes the universe and marks the beginning of the hot Big Bang — has remained unexplored in this quantum gravitational framework. This paper presents the first comprehensive integration of the (generalized) reheating mechanism into the LQC paradigm. Using the Power Law Plateau potential and comparing predictions with the latest Planck 2018 and ACT 2025 data, we demonstrate that the inclusion of a reheating phase with a generic equation of state is fully consistent with the cosmological constraints. In addition, using the observational data for the amplitude and spectral index of the scalar perturbations and the tensor-to-scalar ratio, we also constrain the total number of e-folds from the bounce to the present day and find a lower bound, which is less constrained than that obtained previously from the fitting of  the high-$l$ CMB temperature
power spectrum (TT), the polarization data (TT,TE, EE) and the low-$l$ polarization data (lowP).

\end{abstract}

\maketitle

%%%%%%%%%%%%%%%%%%%%%%%%%%%%%%%%%%%%%%%%%%%%%%%%%%%%%%%%%%%%%%%%%%%%%%%%%%%%%%%%%%%%
\section{Introduction}
\label{introduction}
\renewcommand{\theequation}{1.\arabic{equation}} \setcounter{equation}{0}
%%%%%%%%%%%%%%%%%%%%%%%%%%%%%%%%%%%%%%%%%%%%%%%%%%%%%%%%%%%%%%%%%%%%%%%%%%%%%%%%%%%%%%%%%%%%%
The inflationary scenario is a unique and convincing paradigm that serves as one of the cornerstones of contemporary cosmology. A brief phase of rapid expansion that occurred at the initial instant of the universe, just before the radiation-dominated epoch is referred to as cosmological inflation. At first, inflation was used to solve the shortcomings of the Big Bang models, including the horizon and flatness problems \cite{Guth:1980zm, Starobinsky:1980te,Baumann:2009ds,Linde:1981mu,Linde:1983gd,Guth:1982ec}.
 A successful inflationary model should predict a quasi-adiabatic and almost Gaussian primordial spectrum of perturbations, with spectral properties consistent with the Cosmic Microwave Background (CMB) observations \cite{Komatsu_2011,Planck:2013jfk,Planck:2015sxf,Planck:2018jri,BICEP:2021xfz,ACT:2025fju,ACT:2025tim}. However, despite the fact that inflation is  an elegant paradigm and solves various problems of the big bang universe, it fails to address  other fundamental issues of the early universe, such as  
 the Big Bang singularity and dynamics before the preinflationary phase. In particular, the success of the inflationary scenario depends heavily on our ability to comprehend the ultraviolet (UV) physics. For inflationary models which require the number of e-folds more than $70$ it becomes doubtful whether the underlying quantum field theory on a classical spacetime is reliable or not \cite{Martin:2013tda}.  This is because these theories treat spacetime as classical when the size of the current universe is smaller than the Planck size at the beginning of inflation and this gives rise to the trans-Planckian problem \cite{Brandenberger:2012aj,Martin:2000xs}. Furthermore, the initial singularity is still inevitable \cite{Borde:2001nh,Borde:1993xh}. On the other hand, loop quantum cosmology (LQC) presents a compelling explanation for preinflationary physics which can resolve these issues. 
 In LQC the singularity at Big Bang is replaced by a quantum bounce where the universe evolves  from a contracting phase to an expanding phase smoothly without  hitting any singularity \cite{Ashtekar:2011ni,Bojowald:2015iga,Li:2023dwy}. LQC now becomes a well-established branch of Loop Quantum Gravity (LQG), a quantum gravity theory created to solve the issues of combining quantum mechanics with general relativity (GR), and 
has the potential to act as a quantum gravity extension for a range of cosmological models, including inflation, ekpyrotic, and  matter bounces \cite{Ashtekar:2009mb,Taveras:2008ke,Bojowald:2005cw,Bojowald:2009zzc,Agullo:2012sh,Agullo:2012fc,Agullo:2013ai,Iteanu:2022zha,ElizagaNavascues:2020uyf,ElizagaNavascues:2017avq,CastelloGomar:2017kbo,Fernandez-Mendez:2012poe,Bojowald:2008jv,Mielczarek:2011ph,Cailleteau:2011kr,Cailleteau:2012fy,Cailleteau:2013kqa,Barrau:2014maa,Bojowald:2011hd,Bojowald:2011iq,Zhu:2015xsa,Zhu:2015owa,Zhu:2015ata,Bonga:2015kaa,Bonga:2015xna,Bojowald:2001xe,Ashtekar:2006rx,Ashtekar:2006uz,Ashtekar:2006wn,Ashtekar:2007em,Yang:2009fp,Li:2020pww,Li:2023dwy,Li:2020mfi,Li:2019qzr,Li:2018fco,Li:2018opr,Li:2023dwy,Li:2019ipm,Zhu:2017jew,Zhu:2016dkn,Zhu:2015ata,Li:2018vzr,Qiao:2018dpp,Agullo2023,Levy:2024naz,Barboza:2022hng,Barboza:2020jux,Mohammadi:2024bye,Graef:2018ulg,Graef:2020qwe}. In the conventional inflationary paradigm, the evolutionary trajectory usually initiates well after the Planck era, marked by curvature and energy density of matter fields in the universe approximately a several orders of magnitude lower than the Planck scale, thereby making quantum gravity effects negligible. It is unclear how these pre-inflationary dynamics will be once quantum gravity effects become important. The lack of knowledge regarding earlier cosmic stages is embedded in the selection of initial conditions at the onset of inflation, encompassing both the background homogeneous geometry and cosmological perturbations. The latter is commonly assumed to adhere to the Bunch-Davies vacuum state at the onset of inflation. This assumption is crucial, albeit robust. There is considerable interest in extending this scenario backward in time to encompass the Planck era and demonstrating that these initial conditions (or something akin to them) can arise from the pre-inflationary dynamics when quantum gravity effects become significant. Additionally, in standard inflationary models, the background spacetime exhibits classical singularity, an issue addressed in LQC where quantum gravity effects substitute the big-bang singularity with a non-singular bounce. Similarly, both the ekpyrotic and matter bounce scenarios necessitate a cosmic bounce (See, for example, Refs. \cite{Gasperini:1992em,Gasperini:1992em,Khoury:2001wf,Battefeld:2014uga,Brandenberger:2016vhg,Ijjas:2019pyf,Ijjas:2024oqn}, and references therein), a requirement for which LQC offers a natural mechanism. LQC serves as a quantum gravity culmination for these cosmological scenarios, emphasizing the additional aspects introduced into observable quantities through this extension. The novel effects arising from quantum gravity act as a gateway to the cosmic Planck era, offering an opportunity to test the concepts outlined here by comparing predictions with observations of CMB. Thus, LQC provides a complete description of the spacetime geometry of the Planck era for the Friedman-Lemaitre-Robertson-Walker (FLRW) spacetime. An important question to be addressed is whether the quantum bounce and associated pre-inflationary physics can leave some interesting imprints on the current or future cosmological experiments \cite{Agullo:2016tjh,Ashtekar:2017iip}. There are primarily several distinct methods for studying pre-inflationary dynamics and cosmological perturbations \cite{Agullo2023}. All of these approaches lead to the same set of dynamical equations for the evolution of the background. Therefore, the results we present in this article will be valid in all these approaches. To address the issues of pre-inflationary physics various inflationary models have been considered in the framework of LQC \cite{Ashtekar:2011ni,Li:2023dwy}, and it was shown explicitly that the slow-roll inflation is generic \cite{Ashtekar:2009mm,Ashtekar:2011rm} \footnote{This is also true  in modified LQCs \cite{Li:2019ipm}. For the latter, we refer readers to \cite{Li:2021mop,Saeed:2024xhk} and references therein.}. In this work, we shall conduct a thorough analysis of the impact of the quantum bounce and the pre-inflationary dynamics on the background evolution. One of the main objectives of this work is to study the effect of generalized reheating to LQC.

The paper is organized as follows. In Section~\ref{bckgrnd}, we provide a concise overview of background evolution in LQC, employing the well-motivated Power Law Plateau (PLP) potential and examining its dynamics through both analytical and numerical methods. Section~\ref{inflationary_epoch} is dedicated to a detailed analysis of the PLP model within this framework. In particular, in Section~\ref{inflationary_epoch}.A, using a new method proposed in Appendix A.C, we calculate the six physical quantities ($a_*, H_*, \phi_*, \dot{\phi}_*, k_*, V_0$) at the horizon crossing $k_* = a_*H_*$ for the pivot mode $k_*/a_0 = 0.05/\text{Mpc}$ for the given PLP potential $V(\phi, V_0)$ and observational data $(A_s, n_s)$, where $k_*$ denotes the comoving wavenumber of the pivot mode, and $a_0$ is the current value of the expansion factor of the Universe. With such obtained physical quantities, in Section~\ref{inflationary_epoch}.B we fit our model with the current observational data \cite{Planck:2018jri,BICEP:2021xfz,ACT:2025fju,ACT:2025tim} and show that the model is consistent with observations. In Section~\ref{reh}, we integrate the (generalized) reheating mechanism into the LQC paradigm, and demonstrate that the inclusion of a reheating phase with a generic equation of state is fully consistent with the current cosmological constraints. Section~\ref{rehexp} explores the implications of generalized reheating for the total number of e-folds the Universe has undergone from the quantum bounce to the current time. We conclude with a summary of our findings and future research directions in Section~\ref{conc}. 

Additional technical details and derivations are provided in Appendix A,
in which we first study the quantum effects of LQC to the power spectra, including the amplitudes, spectral indices, running of both scalar and tensor perturbations. Then, we consider the slow-roll conditions in the subsection A.A, and the six independent algebraic equations held at the horizon crossing in the subsection A.B. In the subsection A.C, we propose a method to calculate the six physical quantities ($a_*, H_*, \phi_*, \dot{\phi}_*, k_*, V_0$) at the horizon crossing, using the above six independent algebraic equations. These quantities uniquely determine the evolution of the background universe from the pre-bounce phase down to the onset of the hot Big Bang universe through the dynamical equations of LQC.

%%%%%%%%%%%%%%%%%%%%%%%%%%%%%%%%%%%%%%%%%%%%%%%%%%%%%%%%%%%%%%%%%%%%%%%%%%%%%%%%%%%%%%%%%%%%%%%%%%%%%%%%%%%%%%%%%%%%%%%%%%%%%%%%%%%%%
\section{Background Evolution}
\label{bckgrnd}
\renewcommand{\theequation}{2.\arabic{equation}} \setcounter{equation}{0}
%%%%%%%%%%%%%%%%%%%%%%%%%%%%%%%%%%%%%%%%%%%%%%%%%%%%%%%%
The modified Friedmann equation in the LQC can be written as \cite{Ashtekar:2011ni,Li:2023dwy}  
\bqn
\lb{friedmann}
H^2=\frac{8\pi}{3m_{\text{Pl}}^2}\rho\left(1-\frac{\rho}{\rho_\text{c}}\right),
\eqn
where $H\equiv \dot a/a$ represents the Hubble parameter and the dot stands for the derivative concerning the cosmic time $t$. We also use $m_\text{Pl}=1/\sqrt{G}$ for the Planck mass, and 
$\rho$ and $\rho_\text{c}$ denote the energy density and critical energy density respectively. The maximal value of the critical energy density is  $\rho_\text{c} \simeq 0.41 m_\text{Pl}^{4}$. The occurrence of a non-singular quantum bounce, which eliminated the initial singularity in the early stages of the classical universe, is a remarkable prediction of LQC.   When $\rho=\rho_\text{c}$, the Hubble parameter becomes zero, and energy density takes the maximum value ($\rho(t_B) = \rho_\text{c}$), so the quantum bounce becomes unalterable. Extensive studies have been conducted on the background evolution with a bounce phase \cite{Bonga:2015xna,Zhu:2017jew,Shahalam:2017wba,Shahalam:2018rby,Sharma:2018vnv,Sharma:2019okc,Shahalam:2019mpw,Levy:2024naz}. One of the striking features of the quantum bounce is that the desired slow-roll inflation phase is inevitable \cite{Ashtekar:2009mm,Ashtekar:2011rm} (also see \cite{Ashtekar:2016wpi,Bojowald:2008jv,Mielczarek:2011ph,Cailleteau:2011kr,Cailleteau:2012fy,Cailleteau:2013kqa,Barrau:2014maa,Ashtekar:2011ni, Singh:2006im, Mielczarek:2010bh,Zhang:2007bi,Chen:2015yua}).

In the FLRW background, for a  potential $V(\phi)$ and single scalar field $\phi$ the equation of motion in LQC takes the usual form of Klein-Gordon equation as in GR, given by
\bqn
\lb{klein-gordon}
\ddot \phi +3 H \dot \phi +V_{,\phi}=0,
\eqn
where $V_{,\phi}= dV(\phi)/d\phi$.

 Here, we will study the ``pre-inflationary" and ``slow-roll inflationary" scenarios for the PLP potential, which has the following form \cite{Dimopoulos:2016zhy}
\begin{equation}
V{(\phi)}= V_0 m_\text{Pl}^4 \left( \frac{\phi^2}{ \phi^2+m^2 } \right)~,
\label{potential}
\end{equation}
where $\phi$ is the inflaton field and $V_0$ is the scale of the inflation which can be fixed by matching the amplitude of the scalar power spectrum at the pivot scale of the CMB observations. It is clear that, if $m$ is super-Planckian, then this model becomes indistinguishable from the large field $\phi^n$ models. Thus, in our analysis, we consider $m = m_{\text{Pl}}$, whereas $V_0$ is fixed from the recent observation of CMB (elaborately discussed in Section.\ref{inflationary_epoch}).   

First, we study the background dynamics for inflationary potential given by Eq.~(\ref{potential}). By imposing the initial conditions on  $a(t)$, $\phi (t)$, and $\dot \phi(t)$ at a specific point, one can solve   Eq.~(\ref{friedmann}) and Eq.~(\ref{klein-gordon}) numerically. We set the initial conditions at the bounce $t=t_\text{B}$, using
\bqn
\frac{1}{2} \dot \phi^2(t_{\text{B}})+V(\phi(t_{\text{B}}))=\rho_\text{c}, \;\;\text{and}\;\;\; \dot a(t_{\text{B}})=0.
\label{phid_v}
\eqn
From now on, the subscript `B' represents the variable values at the point of the bounce.
Considering  $\rho_c$ to be a constant and following Eq.~(\ref{phid_v}), one can write $\dot{\phi}_B$  in terms of $\rho_c$ and $\phi_B$ for a given potential. Thus, one needs to choose the specific initial conditions for $a(t_{\text{B}})$, $\phi(t_\text{B})$ and the sign of $\dot{\phi}_B$ only. For the sake of simplicity, we choose $a(t)$ by setting the scale factor at bounce, $a(t_\text{B})=1$. Then, the set of initial conditions  reduces to only the value of $\phi(t_\text{B})$ and  the sign of $\dot{\phi}_B$. In this analysis, both $\dot \phi_\text{B}>0$ and  $\dot \phi_\text{B}<0$ cases are considered.

 To study the pre-inflationary and inflationary dynamics for the given potential, we first introduce the following background quantities:
 \begin{itemize}

\item The equation of state ($\omega(\phi)$), defined as
\bqn
\lb{EOS}
\omega(\phi)\equiv \frac{\dot \phi^2/2-V(\phi)}{\dot \phi^2/2+V(\phi)}.
\eqn
During the slow-roll inflation,  one requires $\omega(\phi) \simeq -1$.

\item The  slow-roll parameters, defined as
\bqn
\lb{epsilonH}
\epsilon_{H}\equiv -\frac{\dot H}{H^2}, \quad \eta_{H}\equiv \frac{\ddot H}{2\dot H H}.
\eqn
Then, we have 
\bqn
\lb{ddotA}
\frac{\ddot{a}}{a} = H^2\left( 1- \epsilon_{H}\right)
= \begin{cases}
> 0, &    \epsilon_{H} < 1, \cr 
= 0, &    \epsilon_{H} = 1, \cr 
< 0, &    \epsilon_{H} > 1. \cr 
\end{cases}
\eqn
During the slow-roll inflation, both $\epsilon_H$ and  $\eta_{H}$ are required to be very small. In standard inflationary scenarios to satisfy the current observation bound on tensor to scalar ratio, $\epsilon_H$  is  required $\epsilon_H \lesssim 10^{-3}$.

\item The duration of the slow-roll inflation is given in terms of the number of $e$-folds ($N_\text{inf}$), which represents the amount of the expansion the universe goes from the onset of inflation to the end of it. It is defined as
\bqn
N_\text{inf}\equiv \ln\left(\frac{a_\text{\text{end}}}{a_i}\right).
\eqn
 When $\ddot a(t_i) = 0$ (or $\epsilon_H(t_i) = 1$), that is, when $\ddot a(t)$ first changes its sign right after the bouncing phase, it defines the onset of the inflation, whereas 
 the inflationary phase ends at $t = t_{\text{\text{end}}}$, here $t_{\text{\text{end}}}$ denotes the first time when 
 \bqn
 \lb{eq2.9}
 \epsilon_H(t_{\text{\text{end}}}) = 1,
 \eqn
 again  after $t_i$, so that $\ddot{a} > 0$
 %\; (\text{and}\; )$ 
 for $t \in (t_i, t_{\text{\text{end}}})$.

\end{itemize}
%%%%%%%%%%%%%%%%%%%%%%%%%%%%%%%%%%%%%%%%%%%%%%%%%%%%%%%%%%%%%%%%%%%%%%%%%%%%%%%%%%%%%%%%%%%%%%%%%%%%%%%%%%%%%%%%%%%%%%%%%%%%%%%%%%%
\subsection{Analytical Evolution of the Background}
%%%%%%%%%%%%%%%%%%%%%%%%%%%%%%%%%%%%%%%%%%%%%%%%%%%%%%%%%%%%%%%%%%%%%%%%%%%%%%%%%%%%%

Background evolution in LQC can be divided universally into three different phases: bouncing, transition, and slow-roll inflationary phase \cite{Zhu:2017jew,Shahalam:2017wba,Shahalam:2018rby,Sharma:2018vnv,Sharma:2019okc,Shahalam:2019mpw}. First, we will present the analytic solutions of the scale factor $(a(t))$ near the bouncing phase. Near the bounce phase, if the kinetic energy is dominant, thus the potential term can be ignored, then Eq.~(\ref{friedmann}) can be rewritten as \cite{Zhu:2017jew}
\bqn
\lb{fri}H^2 =\frac{8\pi}{3m_{\text{Pl}}^2} \frac{1}{2}\dot \phi^2 \left(1-\frac{\dot \phi^2}{2\rho_{\text{c}}}\right),
\eqn
while the field evolution Eq.~(\ref{klein-gordon}) now becomes
\bqn
\lb{kg}\ddot \phi+3 H \dot \phi =0.
\eqn
Solving the above two equations analytically gives
\bqn\lb{phi_dotA}
\dot \phi(t)= \pm \sqrt{2\rho_{\text{c}}} \left(\frac{a_\text{B}}{a(t)}\right)^{3}.
\eqn
Finally, using Eq. (\ref{fri}) and  (\ref{phi_dotA}), the expression of the  scale factor$(a(t))$ can be calculated to be
\bqn\lb{scalar_analytical}
a(t)=a_\text{B}\left(1+\gamma_\text{B} \frac{t^2}{t_{\text{Pl}}^2}\right)^{1/6},
\eqn
where $\gamma_\text{B}\equiv {24\pi \rho_{\text{c}}}/{m_{\text{Pl}}^4} \simeq 30.9$ is a dimensionless constant.

\begin{figure}
\subfigure[\label{atm1phidpos}]{\includegraphics[width=0.49\textwidth]{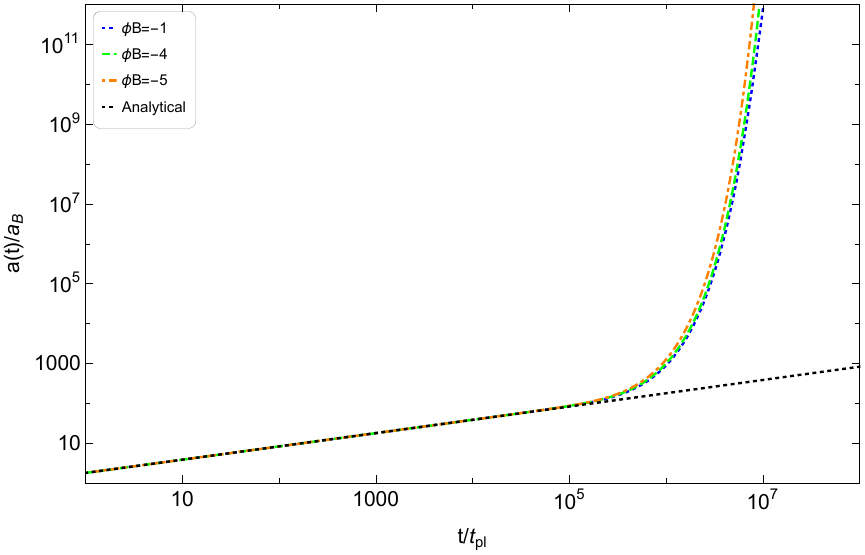}}
\subfigure[\label{wtm1phidpos}]{\includegraphics[width=0.49\textwidth]{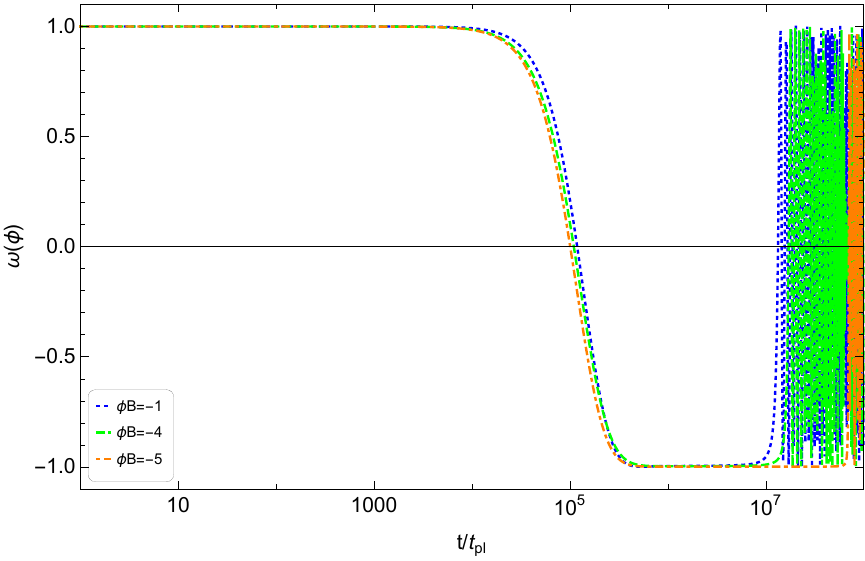}}
\subfigure[\label{eHm1phidpos}]{\includegraphics[width=0.49\textwidth]{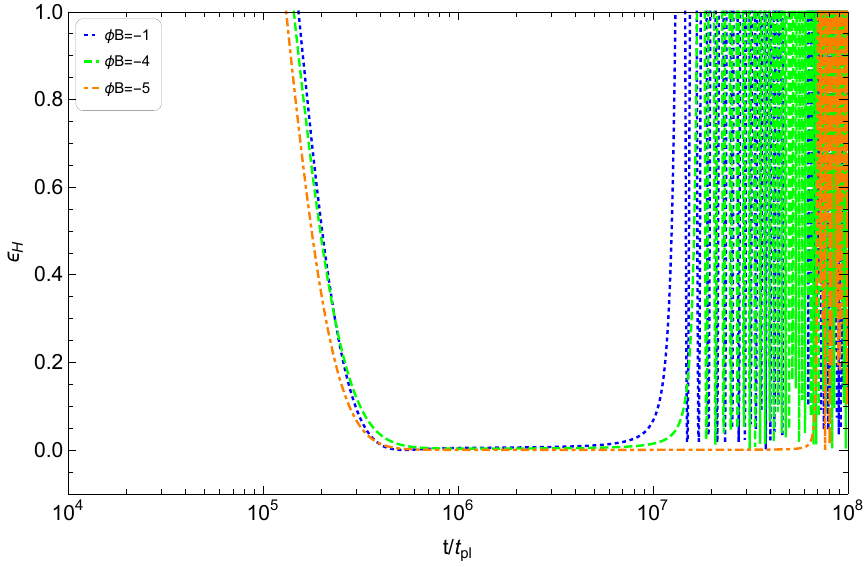}}
\subfigure[\label{energym1phidpos}]{\includegraphics[width=0.49\textwidth]{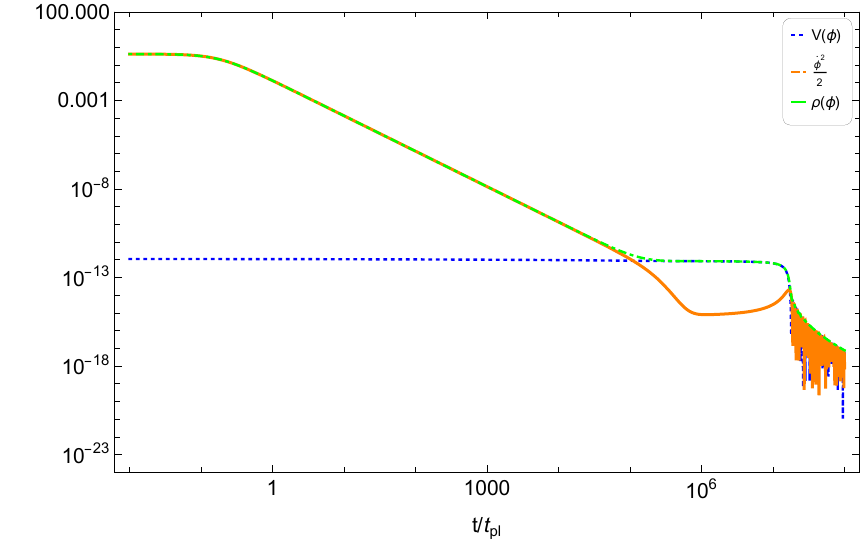}}
\caption{Numerical Evolution of the background quantities for the potential (\ref{potential}), for the kinetic energy dominated at bounce with $\dot \phi>0$. Fig. \ref{atm1phidpos} represents the evolution of the scale factor  $a(t)$ for the different initial values of the field $ \phi$ at the bounce. The black dotted line is the representative of the analytical solution. Fig. \ref{wtm1phidpos} shows the evolution of the equation of state $(\omega(\phi))$ parameter for the different initial values of the $ \phi$. In Fig. \ref{eHm1phidpos} we show the solution for the first slow roll parameter $(\epsilon_{H})$ and in Fig. \ref{energym1phidpos} shows the comparison between the potential $V(\phi)$, kinetic energy density $\dot \phi^2 /2$ along with the total energy density  $\rho = \dot \phi ^2 /2 + V(\phi)$, here we take the initial field value to be $\phi_B=-4$ keeping  $\dot \phi>0$.}
\label{plot1}
\end{figure}
%%%%%%%%%%%%%%%%%%%%%%%%%%%%%%%%%%%%%%%%%%%%%%%%%%%%%%%%%%%%%%%%%%%%%%%%%%%%

\begin{figure}
\subfigure[\label{Nphidneg}]{\includegraphics[width=0.47\textwidth]{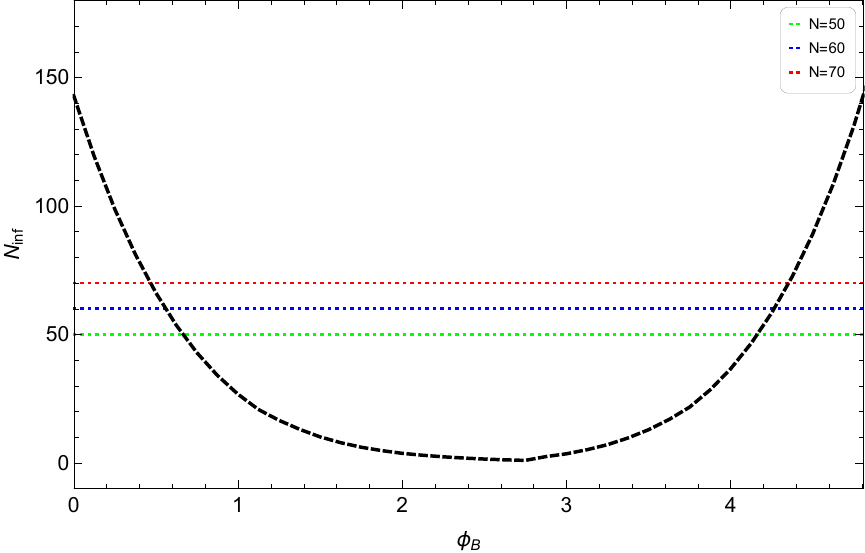}}
\subfigure[\label{Nphidpos}]{\includegraphics[width=0.47\textwidth]{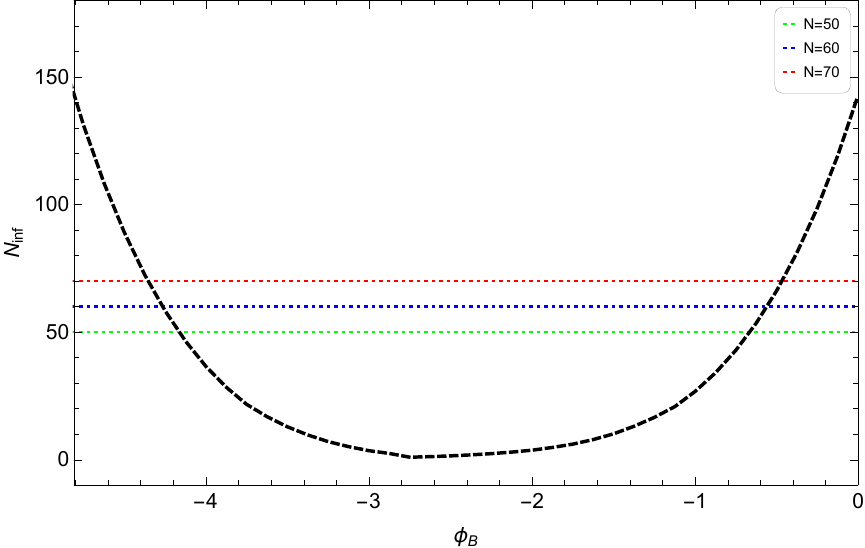}}
\caption{Evolution of e-folds for the slow roll inflation against the different choices of $\phi_B$. In the left panel, we have considered $\dot \phi<0$ whereas in the right panel, we consider  $\dot \phi>0$. Three horizontal dotted lines green, blue, and red denote the $N_{\inf}=50,60,$ and $70$ respectively.}
\label{plot3}
\end{figure}

%%%%%%%%%%%%%%%%%%%%%%%%%%%%%%%%%%%%%%%%%%%%%%%%%%%%%%%%%%%%%%%%%%%%%%%%%%%%%%%%%%%%%%%%%%%%%%%%%%%%%%%%%%%%%%%%%%%%%%
\subsection{Numerical evolution of the Background}
\label{numeicalbackground}
%%%%%%%%%%%%%%%%%%%%%%%%%%%%%%%%%%%%%
In this section, we will present the numerical evolution of the background quantities for the different initial field values ($\phi_B$). As already mentioned we are setting initial conditions at the time of bounce. The difference in the analytical and numerical evolution can be seen clearly in Fig.~\ref{atm1phidpos}. We carried out the numerical evolution of the background quantities for two different cases, 
  (1)  $\phi_B<0$  and $\dot \phi>0$, and 
    (2) $\phi_B>0$  and $\dot \phi<0$.
Interestingly, due to the intrinsic symmetry of the PLP potential, both cases end up with similar results.

\begin{figure}
\subfigure[\label{atm1phidneg}]{\includegraphics[width=0.49\textwidth]{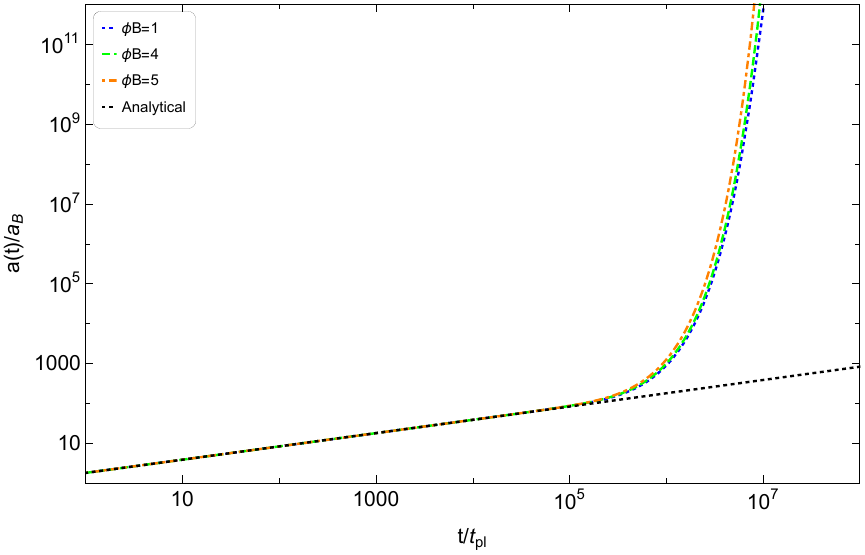}}
\subfigure[\label{wtm1phidneg}]{\includegraphics[width=0.49\textwidth]{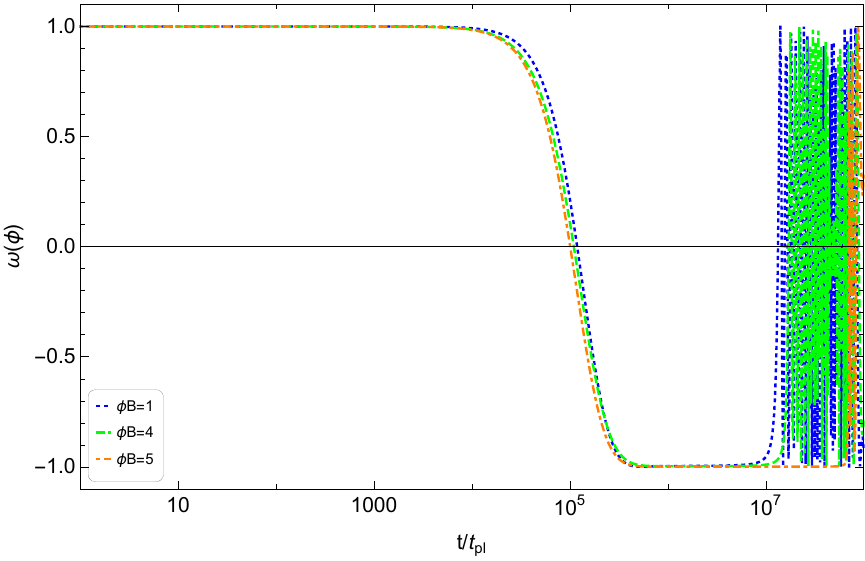}}
\subfigure[\label{eHm1phidneg}]{\includegraphics[width=0.49\textwidth]{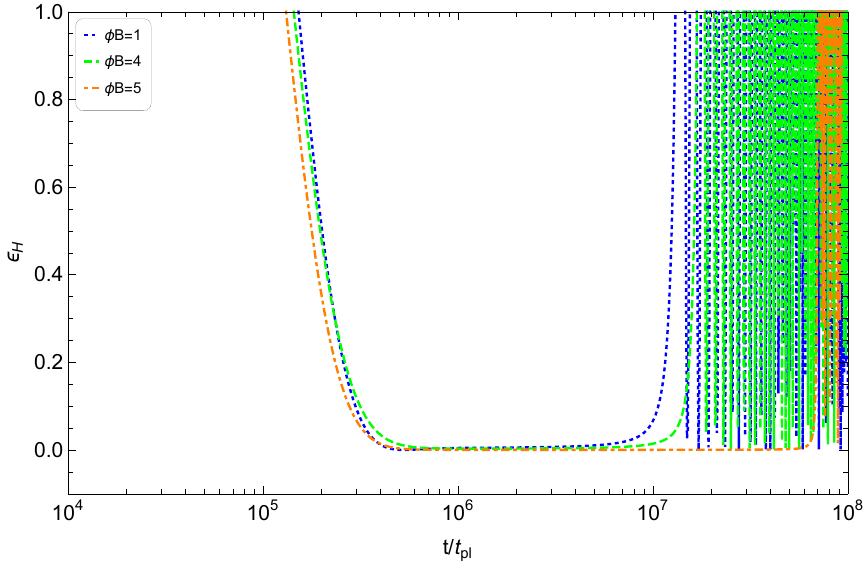}}
\subfigure[\label{energym1phidneg}]{\includegraphics[width=0.49\textwidth]{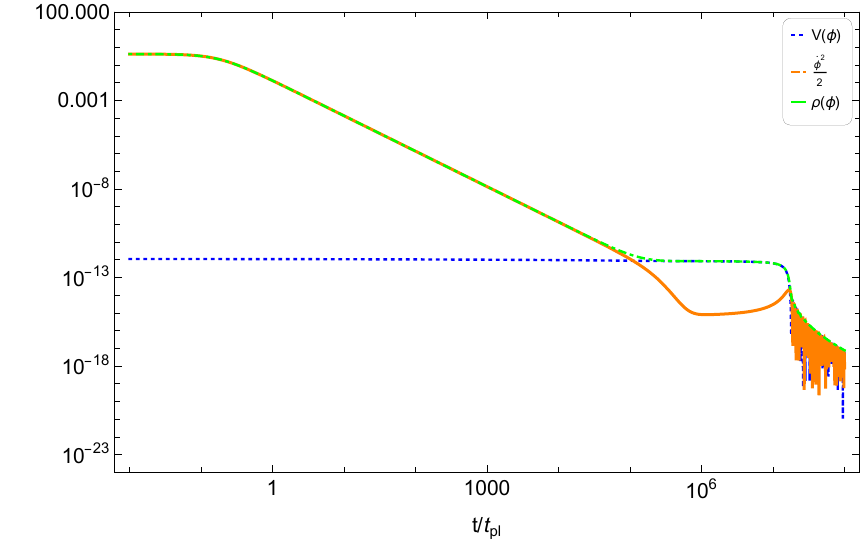}}
\caption{Numerical Evolution of the background quantities for the potential (\ref{potential}), for the kinetic energy dominated at bounce with $\dot \phi<0$. Fig \ref{atm1phidneg} represents the evolution of the scale factor  $a(t)$ for the different initial values of the field $ \phi$. The black dotted is for the analytical solution. Fig \ref{wtm1phidneg} shows the evolution of the equation of state $(\omega(\phi))$ parameter for the different initial values of the $ \phi$. In fig \ref{eHm1phidneg} we show the solution for the first slow roll parameter $(\epsilon_{H})$ and in fig \ref{energym1phidneg} shows the comparison between the potential $V(\phi)$, kinetic energy density $\dot \phi^2 /2$ along with the total energy density  $\rho = \dot \phi ^2 /2 + V(\phi)$, here we take the initial filed value to be $\phi_B=4$ keeping  $\dot \phi<0$.}
\label{plot2}
\end{figure}

The numerical evolution shown in Fig. \ref{plot1} for $\dot\phi_B > 0$ and different background quantities allows us to extend our understanding of the three different phases: bouncing, transition, and slow roll inflation. From Fig.~\ref{atm1phidpos} and \ref{wtm1phidpos} one can see the behavior of the scale factor and the equation of state parameter from the time of the bounce to the inflation. The exponential expansion required for the slow roll inflation is explicitly demonstrated. Around the bounce point, kinetic energy  dominates, leading to the equation of state, $\omega=1$. However, once the kinetic-energy dominates at the bounce, it will dominate  for a long  time [$t \in (0, 10^{5}\; t_{Pl}$)] and after that, it starts to decrease dramatically and quickly reaches the point where  $\omega=-1$, whereby the slow-roll inflation begins. The end of the inflationary phase is achieved when  $\omega \simeq -1/3$ at which we have $\epsilon_H \simeq 1$, as shown in Fig. \ref{eHm1phidneg}. The brief period for  $\omega$ from $1$ to $-1$ is known as the transition phase. On the other hand, from Fig. \ref{energym1phidneg} we can see that the potential energy $V(\phi)$ remains almost constant during the bouncing, transition and slow-roll inflation phases. The slow-roll inflation does not start until the kinetic energy of the inflaton decreases below the potential energy. After a  period of the rapidly exponential expansion, the potential energy drops dramatically, and the inflation ends when it is about equal to the kinetic energy of the inflaton. The e-folds of the expansion of the universe during this epoch depends on the initial conditions $\phi_B$, as shown in Fig. \ref{plot3}.

In Fig.~\ref{plot2}, we show the numerical solution of the background quantities for $\dot \phi<0$. One can notice that the results are similar to the case with $\dot \phi> 0$. This is attributed to the symmetry in the potential around the minima. A common feature that can be noticed from Figs.~\ref{plot1} - \ref{plot2} is that the potential energy remains almost constant during the bounce, transition, and inflationary phases. On the other hand, once at the bounce the evolution of the universe is dominated by kinetic energy, it will dominate the evolution for a long period. Then, at $t \approx 10^{5}\; t_{\text{Pl}}$, it  drops dramatically, whereby the slow roll inflationary epoch starts.

\section{The slow-roll inflationary epoch}
\label{inflationary_epoch}
\renewcommand{\theequation}{3.\arabic{equation}} \setcounter{equation}{0}

After the bounce followed by the transition period, the slow roll inflationary phase begins as $ \omega \approx-1$. In this period the potential energy starts to dominate over the kinetic energy. At this time, sufficiently far away from the bounce, all LQC effects become negligible and the dynamical equations reduce to those given in GR 
\cite{Liddle:1994dx,Baumann:2009ds}
\bqn
H^2 \simeq \frac{8\pi}{3m_\text{Pl}^2} V(\phi),\\
3 H \dot \phi +\frac{dV(\phi)}{d\phi} \simeq 0,
\eqn
with the assumptions  $\frac{1}{2}\dot \phi^2 \ll V(\phi)$ and $|\ddot \phi| \ll |H\dot \phi|$. Using the Friedmann equation for the potential $V(\phi)$ one gets
\bqn\lb{scalar_slow}
a(t)  \propto e^{H_{\text{inf}} t },
\eqn
where $H_\text{inf}$ represents the Hubble parameter during the slow-roll inflation.
Introducing the two parameters $\epsilon_V$ and $\eta_V$ via the relations \cite{Liddle:1994dx,Baumann:2009ds,Riotto:2002yw}
\bqn
\lb{epsilonV}
\epsilon_V= \frac{m_\text{Pl}^2}{16 \pi} \left(\frac{V_{,\phi}(\phi)}{V(\phi)}\right)^2, \; \; \; \; \; \eta_V = \frac{m_\text{Pl}^2}{8 \pi} \left(\frac{V_{,\phi\phi}(\phi)}{V(\phi)}\right),
\eqn
as shown in Appendix A,   the slow-roll inflation conditions 
$\epsilon_H, \;\; \left|\eta_H\right| \ll 1$,  
leads to 
\bqn
\lb{eq3.6}
\epsilon_H \simeq \epsilon_V, \quad \eta_H \simeq \eta_V - \epsilon_V.
\eqn
For details, see Appendix A. The explicit expressions for the slow-roll parameters and e-folds for our current model are given (setting $ m_{\text{Pl}}=1$, which will be adopted in the rest of this paper) by
\bqn
\lb{eq3.7}
&& \epsilon_V = \frac{1}{4 \pi \left(\phi +\phi^3 \right)^2} , \; \; \; \; \; \eta_V = \frac{ \left(1-3 \phi^2\right)}{4 \pi \left( \phi+\phi^3 \right)^2}, \; \; \; \; N_\text{inf} \simeq 8\pi \int_{\phi_\text{\text{end}}}^{\phi_i} \frac{1}{2} \phi \left(1+\phi^2\right) d\phi.  
\eqn

\begin{figure}
{\includegraphics[width=0.47\textwidth]{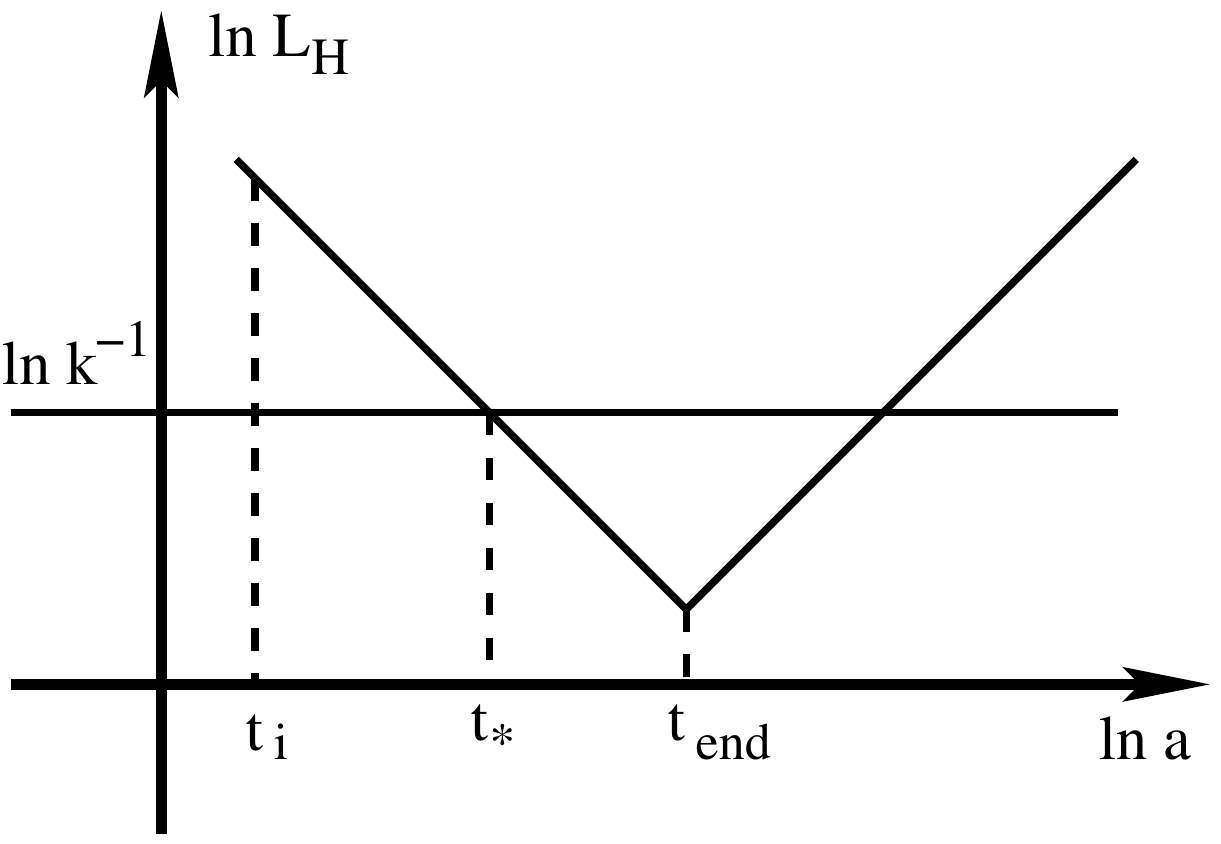}}
\caption{Horizon-crossing (exit) for a given mode $k$ at $t_*$. The line of $\ln L_H \equiv \ln(1/a H)$ is the Hubble horizon. At the exit, we have $k = a(t_*) H(t_*)$.  The pivot mode $k_*$ used by Planck 2018 is $k_*/a_0 = 0.05$/Mpc \cite{Planck:2018jri}.}
\label{fig4}
\end{figure}

\subsection{Calculations of Physical Quantities of the Pivot Mode $k_*$ at Horizon Crossing}

In LQC, using the scaling invariance of the homogeneous and isotropic background of the universe, one often chooses the expansion factor equal to one  at the quantum bounce, that is, $a_B \equiv a(t_B) = 1$, as we did in the last section. Once this is chosen, the
expansion factor can be no longer set to one at the current time, so we cannot identify the comoving modes $k$ to the current observational (physical)  modes $k^{\text{ob}}$. Instead, now   $k^{\text{ob}}$ is given by  $k^{\text{ob}} = k/a_0$, where $a_0$ denotes the current expansion factor and $k$ the comoving wavenumber. In particular, the pivot mode $k^{\text{ob}}_*$ is equal to $ k^{\text{ob}}_* = k_*/a_0$. 
In Appendix A, we present a detailed analysis on how to calculate the power spectra after quantum geometric effects are taken into account and obtain the complete set of equations that the physical quantities must satisfy. One of them is the non-local condition 
\bqn
\lb{eq3.8abbb}
a_B\left(a_*, \phi_*, \dot{\phi}_*\right) = 1,
\eqn
where $\left(a_*, \phi_*, \dot{\phi}_*\right)$ are three of the six physical quantities
\bqn
\lb{eq3.8cbbb}
\left(a_*, \phi_*, \dot{\phi}_*, H_*, k_*, V_0\right),
\eqn
of the pivot mode $k_*^{\text{ob}}$, which are completely determined at the horizon-crossing $k_* = a_*H_*$  through the six algebraic Eqs.(\ref{eqA.38})-(\ref{eqA.42}) and Eq.(\ref{eq3.8abbb}), once $(n_s,A_s)$ are read off from observations. 
It should be noted that once $\left(a_*, \phi_*, \dot{\phi}_*\right)$ are known, the whole evolution of the homogeneous and isotropic universe, from the contracting phase to the end of inflation,  is uniquely determined from the dynamical equations (\ref{friedmann}) and (\ref{klein-gordon}) for a given potential $V(\phi)$.

It must be also noted that for any given mode $k$, there always exists a time $t_*$, at which the mode $k$ crosses the Hubble horizon at the time $t_*$, where $k = a(t_*) H(t_*)$, as shown in Fig. \ref{fig4}. In the rest of this subsection, we shall focus ourselves only on the pivot mode $k_* \;\left(\equiv a_0 k_*^{\text{ob}}\right)$ at the horizon-crossing, where $k_*^{\text{ob}} = 0.05$/Mpc was used by  Planck 2018 and ACT 2025 \cite{Planck:2018jri,ACT:2025fju}. Since $a_0 = a_B e^{N_T}$, where $N_T$ is the total e-fold of the expansion of the Universe from the quantum bounce to the current time, so we have $a_0/a_B =  e^{N_T} \gg 1$. As a result, the comoving pivot mode $k_*$ at the quantum bounce reads  
\bq
\lb{eq3.8bbb}
\frac{k_*}{a_B} = e^{N_T} k_*^{\text{ob}} \gtrsim {\cal{O}}\left(m_{\text{Pl}}\right),
\eq
that is, the pivot mode measured at the quantum bounce is of order of the Planck size, which is expected. 
For the slow-roll inflation, in Appendix A.C we summarize the steps for calculating the six physical quantities listed in Eq.(\ref{eq3.8cbbb}).

\subsection{Fitting the Model with Observational Data}

\begin{figure}
{\includegraphics[width=0.49\textwidth]{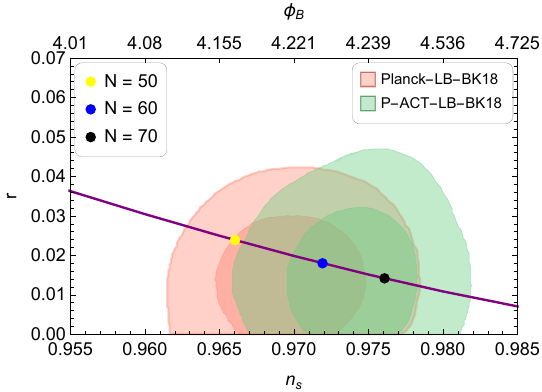}}
\caption{Figure for $(r-n_s)$ with different numbers of e-fold during the inflation, along with the initial field value $\phi_B$ at the bounce. The two light and dark pink (green) contours are for Planck BICEP/Keck Array 2018 (ACT 2025 + Planck BICEP/Keck Array 2018)~\cite{Planck:2018jri,BICEP:2021xfz,ACT:2025fju,ACT:2025tim}, respectively for 1$\sigma$ and  2$\sigma$ bounds. Here the e-fold $N$ is defined as $N \equiv N_* = \ln(a_{\text{end}}/a_*)$, as defined in Eq.(\ref{Ninf_general}).}
\label{fig5}
\end{figure}

To fit the LQC model with observational data, we first note that, in comparing with that of GR, the state at the onset of inflationary is no longer the BD vacuum, due to the particle creations during the pre-inflationary phase. As a result, the parameter $\beta_k$ appearing in the mode function given in Eq.(\ref{eqA.10}) does not vanish \cite{Agullo2023}. Then, the power spectra and indexes of the scalar and tensor modes are given by 
\bqn
 \lb{eq3.13a}
 && P_{\cal{R}}(k) = 2\pi G \left(\frac{  H^2 }{k^3 \epsilon_H}\right) {\cal{F}}(k),  \quad
 P_{h}(k) = 16\pi G  \left(\frac{ H^2 }{k^3}\right) {\cal{F}}(k),
 \eqn
 where \cite{Zhu:2017jew}
 \bqn
 \label{eq3.11a}
  {\cal{F}}(k) \equiv 1 + \delta_{Pl}(k) = 1+ \left[1 + \cos\left(\frac{\pi}{\sqrt{3}}\right)\right]{\text{csch}}^2 \left(\frac{\pi k}{\sqrt{6} k_B}\right), \quad k_B \equiv a_B \sqrt{\frac{8\pi \rho_c}{m_{Pl}^2}}.
 \eqn
Then, introducing the spectral indexes $n_s$ and $n_t$ and their running through Eqs.(\ref{eqA.20}) and (\ref{eqA.21}), we find that
\bqn
 \lb{eq3.13b}
 && n_s(k_*) - 1  =   \xi_{\text{nBD}}(k_*)  - 2\left(3\epsilon_V^* - \eta_V^*\right), \quad 
 n_t(k_*)  \simeq   \xi_{\text{nBD}}(k_*)  -  2\epsilon_V^*, 
 \eqn
where $\xi_{\text{nBD}}(k_*)  \equiv  \left. {d\ln{\cal{F}}(k)}/{d\ln k}\right|_{k = k_*}$, and 
\bqn
  \epsilon_V^* = \frac{1}{4 \pi \left(\phi_*+\phi_*^3 \right)^2} , \; \; \; \; \; \eta_V^* = \frac{ \left(1-3 \phi_*^2\right)}{4 \pi \left( \phi_*+\phi_*^3 \right)^2}.
\label{eq3.12}
\eqn 
 From Planck BICEP/Keck Array 2018 and ACT 2025 observations \cite{Planck:2018jri,BICEP:2021xfz,ACT:2025fju,ACT:2025tim}, we find that  $A_s(k^{ph}_*) \approx 2.09\times 10^{-9}$ for $ k_*/a_0 = 0.05$ Mpc$^{-1}$, which allows us to fix the potential parameter $V_0 \approx 1.63 \times 10^{-12}$. Using this value of $V_0$ along with the relation between $N_{\inf}$ and $\phi_B$ as depicted in Fig. \ref{plot3}, we can establish a relation between $n_s$ and $r_*$, where $r_* = 16 \epsilon_V^*$ as given by Eq.(\ref{eqA.19}) in Appendix A, which holds even after the quantum geometric effects are taken into account.  In  Fig. \ref{fig5} we plot the contours in the ($r, n_s$)-plane with the recent CMB observations of Planck BICEP/Keck Array 2018 and ACT 2025 \cite{Planck:2018jri,BICEP:2021xfz,ACT:2025fju,ACT:2025tim}. From the contours it can be seen that our model is consistent with these observations. In particular, the results with  $N_{*}=50$ are well inside the $1\sigma$ bound, and the ones with $N_{*}=60$ are consistent with the $2\sigma$ bound, where
 $N_* \equiv \ln(a_{\text{end}}/a_*)$ [cf. Eq.(\ref{Ninf_general})].

%%%%%%%%%%%%%%%%%%%%%%%%%%%%%%%%%%%%%%%%%%%%%%%%%%%%%%%%%%%%%%%%%%%%%%%%%%%%%%%%%%%%%%%%%%%%%%%%%%%%%%%%%%%%%%%%%%%%%%%%%%%%%%%%%%%%%%%%%%%%%%%
 \section{Reheating Analysis}
 \label{reh}
 \renewcommand{\theequation}{4.\arabic{equation}} \setcounter{equation}{0}
 %%%%%%%%%%%%%%%%%%%%%%%%%%%%%%%%%%%%%%%%%%%%%%%%%%%%%
 
One of the inevitable epoch to follow the inflationary paradigm is reheating \cite{Dolgov:1982yb,1982PhLB..117...29A,PhysRevLett.48.1437}, where the inflaton field transfers its energy to other degrees of freedom and resurrects the universe from super-cooled state to a hot thermal bath of relativistic particles. 
In the standard inflationary scenario, numerous methods of reheating have been proposed in the literature such as perturbative decay where the inflaton field reaches the bottom of the potential and starts to decay to other elementary particles \cite{Martin:2014nya,Bassett:2005xm,Rehagen:2015zma,Garcia:2020wiy}. The produced particles interact with each other and reach an equilibrium at a temperature known as reheating temperature ($T_{\text{re}}$). There are other theoretically interesting models that were introduced later, such as tachyonic instability \cite{Felder:2001kt} and parametric resonance \cite{Traschen:1990sw,Shtanov:1994ce,Kofman:1994rk,Kofman:1994rk,Kofman:1997yn,Kofman:1997pt}. The latter is  non-perturbative and often referred to as preheating, an initial stage of the reheating \cite{Kofman:1997yn,Kofman:1997pt,Allahverdi:2010xz,Amin:2014eta,Lozanov:2019jxc}. 
Preheating is more efficient as compared to perturbative reheating. In preheating the decay happens exponentially generating the high number of particles. 

\begin{figure}
{\includegraphics[width=0.49\textwidth]{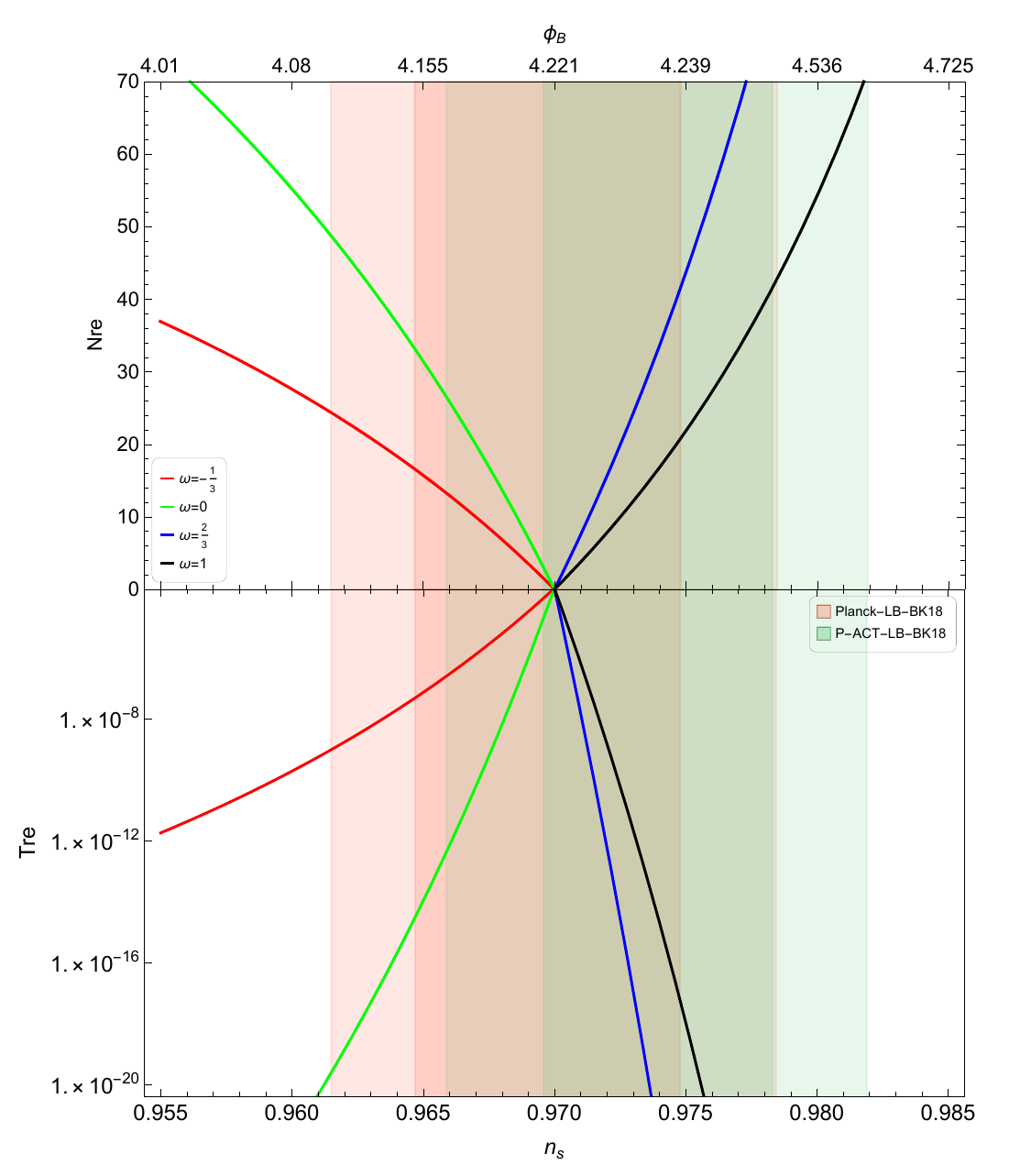}}
\caption{Plot for the number of e-folds $N_{\text{re}}$ and reheating temperature $T_{\text{re}}$ against the $n_s$ ($\phi_B$). The two light and dark pink (green) strips are for Planck BICEP/Keck Array 2018 data (ACT 2025 + Planck BICEP/Keck Array 2018 data)~\cite{Planck:2018jri,BICEP:2021xfz,ACT:2025fju,ACT:2025tim}, respectively for 1$\sigma$ and  2$\sigma$ bounds.}
\label{fig6}
\end{figure}

In \cite{Zhu:2018smk}, using the {\em Uniform Asymptotical Approximation method} \cite{Olver1974AsymptoticsAS,Habib:2002yi,Wang:2010an,Zhu:2013fha,Zhu:2013upa,Zhu:2014wda,Zhu:2014aea,Zhu:2014wfa,Zhu:2015xsa,Zhu:2015owa,Zhu:2015ata,Wu:2017joj,Qiao:2018dpp,Zhu:2018smk,Li:2018vzr,Li:2019cre,Ding:2019nwu,Qiao:2019hkz,Pan:2023aiv,Pan:2025ppa}, the parametric resonance was studied analytically and the   analytical solutions of the mode function were found for some given effective potentials. However, the epoch of reheating can be studied in great detail without delving deep into the microphysical dynamics of this phase \cite{Cook:2015vqa}. As there is no direct observational bound on the reheating temperature, analysis of this era in an indirect approach can be extremely useful. In this approach, one can give a bound on the thermalization temperature through the inflationary observables. Moreover, such approaches can be used as a new way to constrain different inflationary models. Along with the reheating temperature ($T_{\text{re}}$) and the equation of state parameter ($\omega_{\text{re}}$), another important physical quantity is the duration of reheating dubbed as the number of e-folds ($N_{\text{re}}$). $N_{\text{re}}$ quantifies the expansion of the universe from the end of inflation to the end of the thermalization period. Following \cite{Cook:2015vqa, Bhattacharya:2019ryo, Khan:2022odn,Gangopadhyay:2022vgh,Adhikari:2019uaw} reheating temperature ($T_{\text{re}}$) and duration of the reheating ($N_{\text{re}}$) can be written in terms of the effective equation of state ($\omega_{\text{re}}$)  as 
\bqn
\label{Nre}
N_{\text{re}} &=& -\frac{4}{\left(1- 3w_{\text{re}}\right)}\left[\frac{1}{4}\ln\left(\frac{3^2\cdot 5}{\pi^2g_{\text{re}}}\right) + \frac{1}{3}\ln\left(\frac{11g_{\text{re}}}{43}\right)   +\ln\left(\frac{k_*}{a_0T_0}\right) + \ln\left(\frac{V_{\text{end}}^{1/4}}{H_*}\right) + N_*\right],\\
\label{Tre}
T_{\text{re}} &=& \left[ \left(\frac{43}{11 g_{\text{re}}} \right)^{\frac{1}{3}}    \left(\frac{a_0}{k_{*}}\right) T_0 H_{*} e^{- N_{*}} \left(\frac{3^2\cdot 5 V_{\text{end}}}{\pi^2 g_{\text{re}}} \right)^{- \frac{1}{3(1 + w_{\text{re}})}}  \right]^{\frac{3(1+ w_{\text{re}})}{3 w_{\text{re}} -1}}.
\eqn
Here $g_{\text{re}}$ is the relativistic degrees of freedom, 
which is of order   $g_{\text{re}} \simeq {\cal{O}}(100)$ \cite{Cook:2015vqa}, 
and $T_0$ denotes the present value of temperature,  while $k_*$ signifies  {\it Planck's} pivot scale, as noted above, and $H_*$ represents the Hubble parameter measured at the horizon crossing of this pivot mode, and $V_{\text{end}}$ is the potential at the end of inflation. The number of e-folds  $N_*$ is defined as
\bqn\lb{Ninf_general}
N_* &\equiv& \ln\left(\frac{a_\text{\text{end}}}{a_*}\right) = \int_{t_*}^{t_\text{\text{end}}} H(t) dt =  \int_{\phi_*}^{\phi_\text{\text{end}}} \frac{H}{\dot \phi} d\phi \simeq \frac{8\pi}{m_{\rm Pl}^2}\int_{\phi_\text{\text{end}}}^{\phi_*} \frac{V}{V_{,\phi}} d\phi,
\eqn
where  $\phi_{*}$ and $\phi_{\text{end}}$ denote the field values at the horizon crossing  and at the end of inflation, respectively. Note that in writing the last expression of the above equation, we had used the slow-roll condition. In addition, from Eqs.(\ref{eqA.19}) and (\ref{eqA.20}) we find that the Hubble parameter can be written in terms of inflationary observables as 
\begin{equation}
{H_*}= \left(\frac{\pi r_* A_s(k_*)}{16{\cal{F}}(k_*)}\right)^{1/2}\; m_{\text{PL}}.
\label{Hk}
\end{equation}

Then,  maintaining $A_s(k_0)= 2.09\times10^{-9}$ (variation of $A_s$ has a negligible effect on $T_{\text{re}}$ and $N_{\text{re}}$), the rest of the calculations are carried straightforward.  
It is evident from Eq.~(\ref{Nre}) and Eq.~(\ref{Tre}) that both equations depend on $H_*$. From Eq.(\ref{Hk}), it's clear that $H_*$ is a function of the tensor to scalar ratio ($r_*$) and $k_*$. The latter  was calculated in the last section (Section III.A). We write $r_*$ in terms of scalar spectral index ($n_s$) which would help us to constrain $T_{\text{re}}$ and $N_{\text{re}}$ more accurately. From $\epsilon_H=1$ as the end of inflation, we can calculate $V_{\text{end}}$, equipped with all the preliminaries now we can calculate  $T_{\text{re}}$ and $N_{\text{re}}$ for different equations of state ($\omega_{\text{re}}$). The results for the reheating analysis can be found in Fig. \ref{fig6}. We have considered four different values of  ($\omega_{\text{re}}= -1/3,0,2/3,1$), however, one can easily extend the calculations for other values of $\omega_{\text{re}}$. The solid lines with respectively red, green, blue, and black color signify the different $\omega_{\text{re}}$ as mentioned in Fig.~\ref{fig6}. The dark  and light pink (green) shaded strips  represent Planck BICEP/Keck Array 2018 (ACT 2025 + Planck BICEP/Keck Array 2018) data, respectively for 2$\sigma$ and  1$\sigma$ bounds. 
In Fig. \ref{fig6} we have plotted $T_{\text{re}}$ from the scale of instantaneous reheating to the Big Bang Nucleosynthesis (BBN), which is around $10 \text{MeV}$. The point where all the curves in $N_{\text{re}}$ and $T_{\text{re}}$ merge represents the scale of instantaneous reheating  ($N_{\text{re}}=0$). As our inflationary model, PLP,  has an origination from supersymmetric theory, and $g_{\text{re}}$ is taken to be $226$ for the rest of the calculation. From this figure it can be seen clearly that the reheating phase can be smoothly integrated into the LQC framework, and the model is well consistent with current observations.

%%%%%%%%%%%%%%%%%%%%%%%%%%%%%%%%%%%%%%%%%%%%%%%%%%%%%%%%%%%%%%%%%%%%%%%%%%%%%%%%%%%%%%%%%%%%%%%%%%%%%%%%%%%%%%%%%%%%%%%%%%%%%%%%%%%%%%

\section{Reheating and expansion history}
\label{rehexp}
\renewcommand{\theequation}{5.\arabic{equation}} \setcounter{equation}{0}
%%%%%%%%%%%%%%%%%%%%%%%%%%%%%%%

The total e-fold ($N_T$) of the expansion of the universe from the bounce till today  can be expressed in terms of the scale factor in the form
\bqn
N_T = \ln \left( \frac{a_0}{a_B}  \right) = \ln \left( \frac{a_*}{a_B} \cdot \frac{a_{\text{end}}}{a_*} \cdot \frac{a_{\text{re}}}{a_{\text{end}}} \cdot \frac{a_0}{a_{\text{re}}} \right)
=  \ln\left(\frac{a_*}{a_B}\right) + N_* + N_{\text{re}} +  \ln\left(\frac{a_0}{a_{\text{re}}}\right),
\label{nt}
\eqn
where $N_{\text{re}}$ and $N_*$  are given respectively by Eqs.(\ref{Nre}) and (\ref{Ninf_general}), and 
\bqn
\frac{a_0}{a_{\text{re}}} = \left ( \frac{11 g_{\text{re}}}{43}\right)^{\frac{1}{3}} \frac{T_{\text{re}}}{T_0} = \left ( \frac{11 g_{\text{re}}}{43}\right)^{\frac{1}{3}} \left[\left(\frac{43}{11 g_{\text{re}}} \right)^{\frac{1}{3}}\left(\frac{a_0 }{k_{*}}\right) T_0 H_{*} e^{- N_{*}} \left(\frac{3^2\cdot 5\; V_{\text{end}}}{\pi^2 g_{\text{re}}} \right)^{- \frac{1}{3(1 + w_{\text{re}})}}  \right]^{\frac{3(1+ w_{\text{re}})}{3 w_{\text{re}} -1}}\frac{1}{T_0}.
\label{aoare}
\eqn
They are all functions of $H_*$, $\phi_*$ and $\phi_{\text{end}}$ for any given $T_0, \; g_{\text{re}}$ and $w_{\text{re}}$. 
However, $\phi_{\text{end}}$ is fixed by $\epsilon_V(t_{\text{end}}) =1$. Then, in the framework of the slow-roll inflation, at the horizon-crossing $t_*$ we have six equations, Eqs.(\ref{eqA.38b})-(\ref{eqA.44b}),  for the six unknown parameters, $(a_*, H_*, \phi_*, \dot\phi_*, k_*, V_0)$. Based on the analysis carried out in Section III.A and details in Appendix A.C, we can proceed to computing the values of the relevant physical quantities at the horizon crossing, given the forms of the inflationary potentials.

\begin{figure*}[t]
{\includegraphics[width=0.49\textwidth]{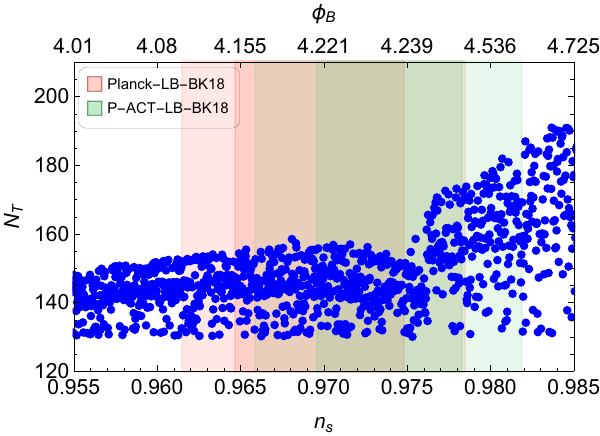}}
\caption{Plot of $N_{\text{T}}$ for the total number of e-folds  against $n_s$. The two light and dark pink (green) strips are for Planck BICEP/Keck Array 2018 data (ACT 2025 + Planck BICEP/Keck Array 2018 data)~\cite{Planck:2018jri,BICEP:2021xfz,ACT:2025fju,ACT:2025tim}, respectively for 1$\sigma$ and  2$\sigma$ bounds.}
\label{fig7}
\end{figure*}

Substituting Eqs. (\ref{aoare}) and (\ref{Nre}) into Eq. (\ref{nt}) one can write the total number $N_T$ of e-folds in terms of ($a_*, H_*, \phi_*, \phi_{\text{end}},\omega_{\text{re}}, g_{\text{re}}, T_0$). It is well known that one needs $N_{*}$ to be in the range of $50-65$, in order to solve the problems of the Big Bang cosmology.    
 In our analysis, we are not assuming $N_{\text{re}}=0$ (instantaneous reheating scenario), rather allowed the reheating to be the general one.  Furthermore, in our analysis $N_{\text{re}}$ and $T_{\text{re}}$ both can vary to a wide range from instantaneous to the BBN temperature depending on the value of $\omega_{\text{re}}$ and $n_s$. This allows us to compute $N_T$ for different values of $\omega_{\text{re}}$ and $n_s$.

In Fig. \ref{fig7}, we plot $N_T$ vs $n_s$, from which we find that  
\bqn
\lb{3.25}
N_T \gtrsim N_T^{\text{min}} \approx 130,
\eqn
which is lower than the bound, $N_T \gtrsim   141$, obtained previously \cite{Zhu:2017jew}. A detailed analysis shows that these results do not contradict. In particular, the low bound obtained in  \cite{Zhu:2017jew} came from the fitting of the model with the high-$l$ CMB temperature
power spectrum (TT), the polarization data (TT,
TE, EE) and the low-$l$ polarization data (lowP) from Planck 2015  \cite{Planck:2015sxf}, while in this paper we mainly consider the  constraints from the current observational data of $(A_s, n_s, r)$ \cite{Planck:2018jri,BICEP:2021xfz,ACT:2025fju,ACT:2025tim}. Therefore, the above different results just show  that  the former imposes more stringent constraints than the latter.

 %%%%%%%%%%%%%%%%%%%%%%%%%%%%%%%%%%%%%%%%%%%%%%%%%%%%%%%%%%%%%%%%%%%%%%%%%%%%%%%%%%%%%%%%%%%%%%%%%%%%%%%%%%%%%%%%%%%%%%%%%%%%%%%%%%%%%%%%%%%%%%%%%%%%%%%
 \section{Conclusion}
 \label{conc}
%%%%%%%%%%%%%%%%%%%%%%%%%%%%%%%%%

This work has presented the first analysis of the reheating epoch within the framework of loop quantum cosmology, thereby completing the cosmic history from the non-singular quantum bounce to the present day. By integrating a generalized reheating phase into the well-established LQC background evolution—comprising the bounce, transition, and slow-roll inflation — we have bridged a critical gap in the literature.
Our key findings and implications are as follows:

\begin{itemize}
    
\item {\em Consistency with Observations:} We have demonstrated that the inclusion of a reheating phase with a generic equation of state is fully consistent with the latest cosmological constraints from Planck 2018 and ACT 2025 data \cite{Planck:2018jri,BICEP:2021xfz,ACT:2025fju,ACT:2025tim}. This establishes the viability of the LQC paradigm through the entire cosmic evolution—from the quantum bounce through inflation and into the hot Big Bang.

\item {\em Constraining Cosmic History:} By utilizing observational data for the amplitude and spectral index of scalar perturbations and the tensor-to-scalar ratio, we have placed a lower bound on the total number of e-folds from the bounce to the present day, finding $N_T \gtrsim 130$.  This bound, derived from the 
($A_s, n_s, r)$ dataset, is less stringent than that obtained from a direct fit to the full CMB power spectrum  \cite{Zhu:2017jew}, highlighting how different observational priors can refine our understanding of the pre-inflationary epoch.

\item{\em A Unified Picture:} This research bridges a critical gap in the LQC narrative. The theory has long provided a robust, singularity-free description of the universe from the quantum bounce through the slow-roll inflation. Our work now extends this coherent picture to include the subsequent transition to a hot, thermal universe via reheating, thereby connecting the quantum gravity regime directly to the standard hot Big Bang cosmology.
 
\end{itemize}

In conclusion,  this study not only reinforces the robustness of LQC as a compelling scenario for the origin of our universe but also opens up new avenues for testing quantum gravitational effects through their imprints on the thermal history of the cosmos \cite{Agullo2023}. Future work will involve extending this analysis to other well-motivated inflationary potentials and loop quantum gravity related models, such as mLQC-I and mLQC-II \cite{Li:2021mop}, and exploring the implications of these findings for the production of primordial black holes and stochastic gravitational wave backgrounds  \cite{Martin:2014nya,Bassett:2005xm,Rehagen:2015zma}.

%%%%%%%%%%%%%%%%%%%%%%%%%%%%%%%%%%%%%%%%%%%%%%%%%%%%%%%%%%%%%%%%%%%%%%%%%%%%%%%%%%%%%%%%%%%%%%%%%%%%%%%%%%%%%%%%%%%%%%%%%%%%%%%%%%%%%%%%%%%%%
\section*{Acknowledgement}
%%%%%%%%%%%%%%%%%%%%%%%%%%%%%%%%%%%%%%%%%%%%%%%%%%%%%%%%%%%%%%%%%%
The authors, Y. and B.F.L., would like to thank Drs. Abolhassan Mohammadi and Tao Zhu for fruitful discussions.  
M.R.G. thanks OMEG Institute, Soongsil University, Republic of Korea, for their hospitality, where the last part of the paper was completed.
A.W. thanks the hospitality of the Centre for Cosmology and Science Popularization, SGT University, India, during his visit (November - December 2024), during which the project was initiated.
Work of M.R.G. is supported by the Department of Science and Technology(DST), Government of India under the Grant Agreement number IF18-PH-228 (INSPIRE Faculty Award) and by the Science and Engineering Research Board (SERB), DST, Government of India under the Grant Agreement number CRG/2022/004120 (Core Research Grant). B.-F. Li is supported by the National Natural Science Foundation of China (NNSFC) grant No. 12005186. A.W. is partially supported by the US NSF grant: PHY-2308845. 
%%%%%%%%%%%%%%%%%%%%%%%%%%%%%%%%%%%%%%%%%%%%%%%%%%%%%%%%%%%%%%%%%%%%%%%%%%%%%%%%%%%%%%%%%%%%%%%%%%%%%%%%%%%%%%%%%%%%%%%%%%%%%%%%%

 %%%%%%%%%%%%%%%%%%%%%%%%%%%%%%%%%%%%%%%%%%%%%%%%%%%%%%%%%%%%%%%%%%%%%%%%%%%%%%%%%%%%%%%%%%%%%%%%%%%%%%%%%%%%%%%%%%%%%%%%%%%%%%%%%%%%%%%%%%%%%%%%%%%%%%%
 \section*{Appendix A: Power Spectra and Slow Roll Parameters with General Initial Conditions}
 \label{AppendixA}
\renewcommand{\theequation}{A.\arabic{equation}} \setcounter{equation}{0}

 In the framework of LQC, the inflationary phase generically happens after the quantum  bounce when the energy density falls to the order of $\rho/\rho_c 
 \simeq 10^{-12}$ \cite{Ashtekar:2011ni,Agullo2023}. In such a scale, the field equations can be well approximated by the classical ones  
 \bqn
 \lb{eqA.1}
 H^2 &\approx& \frac{8\pi G}{3} \rho,\\
  \lb{eqA.2}
  \frac{\ddot{a}}{a} &\approx& - \frac{4\pi G}{3}\left(\rho + 3 P\right),
 \eqn
 where $\rho$ and $P$ denote respectively the energy density and pressure of the perfect fluid, $\dot{a} \equiv d a(t)/dt$, and so on. For a single scalar field $\phi$ with a potential $V(\phi)$ we have
  \bqn
 \lb{eqA.3}
\rho = \frac{1}{2}\dot\phi^2 + V(\phi), \quad P = \frac{1}{2}\dot\phi^2 - V(\phi).
 \eqn
Taking the derivative of Eq.(\ref{eqA.1}) with respect to $t$ and then using Eqs.(\ref{eqA.2}) and (\ref{eqA.3}) we obtain 
  \bqn
 \lb{eqA.4}
\ddot\phi + 3H\dot\phi + V_{,\phi}(\phi) = 0,  
 \eqn
which is nothing but the familiar Klein-Gordon equation. 

To study inflationary universe, two sets of parameters, the so-called slow roll parameters, are often introduced: the one ($\epsilon_H, \eta_H$) defined in terms of the Hubble parameter $H$
and its derivatives, given by Eq.(\ref{epsilonH}), and the one ($\epsilon_V, \eta_V$) defined in terms of the scalar field potential $V(\phi)$ and   its derivatives, given by Eq.(\ref{epsilonH}), which are \footnote{Note that the parameter $\eta_H$ used in this paper is the same as $\delta$ introduced in \cite{Agullo:2013ai} but has a sign difference from $\eta$ introduced in \cite{Baumann:2009ds}, i.e., $\eta_H = \delta = - \eta$.}
 \bqn
\lb{eqA.4b}
\epsilon_{H}\equiv -\frac{\dot H}{H^2}, \quad \eta_{H}\equiv \frac{\ddot H}{2\dot H H}, \quad  
\epsilon_V= \frac{m_\text{Pl}^2}{16 \pi} \left(\frac{V_{,\phi}(\phi)}{V(\phi)}\right)^2, \quad 
\eta_V = \frac{m_\text{Pl}^2}{8 \pi} \left(\frac{V_{,\phi\phi}(\phi)}{V(\phi)}\right).
\eqn
However, the two sets of parameters are connected through the field equations (\ref{eqA.1}) and (\ref{eqA.2}).  To show their relations,  in the following we first consider the general case and then restrict ourselves to the slow-roll case.

To the above purposes, let us first note that 
\bqn
\lb{eqA.5}
\frac{\ddot{a}}{a} = H^2\left(1-\epsilon_H\right) = -4\pi G\dot\phi^2 + H^2,
\eqn
so that we have 
\bqn
\lb{eqA.6}
 \epsilon_H = \frac{4\pi G\dot\phi^2}{H^2},\quad \text{or} \quad \frac{H^2}{\dot\phi^2} = \frac{4\pi G}{\epsilon_H}.
\eqn
In addition, we also have
\bqn
\lb{eqA.7}
\dot{H} = \frac{\ddot{a}}{a} - H^2 =  -4\pi G\dot\phi^2, \quad \ddot{H} = -8\pi G \dot\phi\ddot\phi,
\eqn
so that
\bqn
\lb{eqA.8}
\eta_{H} = \frac{\ddot{H}}{2H\dot{H}}  =  \frac{\ddot\phi}{H\dot\phi} =  \frac{1}{H}\frac{d}{dt}\left(\ln\left|\dot\phi\right|\right).
\eqn

On the other hand, when the background is de Sitter spacetime, both of the scalar and tensor perturbations satisfy the MS equation \cite{Baumann:2009ds}
\bqn
\lb{eqA.9}
\nu_k'' + \left(k^2 - \frac{2}{\eta^2}\right)\nu_k = 0,
\eqn
which has the general solution
\bqn
\lb{eqA.10}
\nu_k = \frac{1}{\sqrt{2k}}\left\{\alpha_k e^{-i k\eta}\left(1 - \frac{i}{k\eta}\right) + \beta_k e^{i k\eta}\left(1 + \frac{i}{k\eta}\right)\right\},
\eqn
where $\alpha_k$ and $\beta_k$ are two integration constants subjected to the Wronskian condition  
\bqn
\lb{eqA.11}
\left|\alpha_k\right|^2 -  \left|\beta_k\right|^2 = 1.  
\eqn
The Bunch-Davies (BD) initial conditions correspond to $\alpha_k = 1$ and $\beta_k = 0$. However, in LQC the initial conditions are usually imposed in the remote contracting phase \cite{Ashtekar:2011ni,Agullo2023}. Therefore, even initially the system is in the vacuum state $\beta_k$ will be different from zero at the onset of the inflation, due to particle creations in the pre-inflationary phase. Then, we find the two-point correlation functions $P_{\cal{R}}(k)$ and $P_h(k)$ are given by
\bqn
\lb{eqA.12}
\left<{\cal{R}}_{\vec k} {\cal{R}}_{\vec{k}'}\right> =    \left(2\pi\right)^3 \delta\left(\vec k + \vec k'\right) \frac{H^2}{\dot\phi^2}\left|\frac{\nu_k}{a}\right|^2 \equiv \left(2\pi\right)^3 \delta\left(\vec k + \vec k'\right) P_{\cal{R}}(k), \\
\lb{eqA.13}
\left<h_{\vec k} h_{\vec{k}'}\right>  =  \left(2\pi\right)^3 \delta\left(\vec k + \vec k'\right) {32\pi G}\left|\frac{\nu_k}{a}\right|^2 \equiv \left(2\pi\right)^3 \delta\left(\vec k + \vec k'\right) P_{h}(k),
\eqn
where the comoving curvature perturbation ${\cal{R}}$ of the scalar modes and the spin-2 tensor perturbation $h_k$ are given respectively by
\bqn
\lb{eqA.14}
{\cal{R}} = \frac{H}{a\dot\phi}\; \nu_k, \quad h_k = \sqrt{16\pi G}\; \frac{\nu_k}{a}.  
\eqn
In the  de Sitter background we have $a = 1/(H\eta)$. Then, from Eq.(\ref{eqA.10}) we find
\bqn
\lb{eqA.15}
\left|\frac{\nu_k}{a}\right|^2 =  \frac{H^2}{2k^3}\Bigg\{\left(\left|\alpha_k\right|^2 +  \left|\beta_k\right|^2\right)\left(k^2\eta^2 +1\right)
+ \alpha_k\beta_k^*e^{-2ik\eta}\left(k^2\eta^2 - 2ik\eta -1\right)  +  \alpha_k^*\beta_ke^{2ik\eta}\left(k^2\eta^2 + 2ik\eta -1\right)\Bigg\}. ~~~~~~~
\eqn 
On the superhorizon scales, we have $|k\eta| \ll 1$ \cite{Baumann:2009ds}, so the above expression reduces to 
\bqn
\lb{eqA.16}
\left|\frac{\nu_k}{a}\right|^2=  \frac{H^2}{2k^3}{\cal{F}}(k), \quad
{\cal{F}}(k) \equiv \left|\alpha_k\right|^2 +  \left|\beta_k\right|^2
- \left(\alpha_k\beta_k^*  +  \alpha_k^*\beta_k\right). ~~~~
\eqn 
Inserting it into Eqs.(\ref{eqA.12}) and (\ref{eqA.13}), we find
\bqn
\lb{eqA.17}
P_{\cal{R}}(k) &=& 2\pi G \left(\frac{  H^2 }{k^3 \epsilon_H}\right) {\cal{F}}(k), \\
\lb{eqA.18}
P_{h}(k) &=& 16\pi G  \left(\frac{ H^2 }{k^3}\right) {\cal{F}}(k).
\eqn
 Note that in writing down Eq.(\ref{eqA.17}) we had used Eq.(\ref{eqA.6}). Then,  we find that
\bqn
\lb{eqA.19}
r \equiv \frac{P_t(k)} {P_{\cal{R}}(k)} =  \frac{2P_h(k)} {P_{\cal{R}}(k)}  = 16 \epsilon_H.  
\eqn

To connect the power spectra to observations, let us introduce the spectrum indices $n_s$ and $n_t$ and their runnings via the relations 
\bqn
\lb{eqA.20}
\Delta^2_{\cal{R}}(k) &\equiv& \frac{k^3}{2\pi^2}P_{\cal{R}}(k) =  \frac{H^2}{\pi m_{PL}^2 \epsilon_H} \; {\cal{F}}(k) \equiv A_s\left(\frac{k_*}{a_0}\right)\left(\frac{k}{k_*}\right)^{n_s(k_*) -1 + \frac{\alpha_s}{2}\ln(k/k_*)
+  \frac{\beta_s}{6}\ln^2(k/k_*) + ...}, \\
\lb{eqA.21}
\Delta^2_{t}(k) &\equiv&  \frac{k^3}{2\pi^2}P_{t}(k)  =  \frac{16 H^2}{\pi m_{PL}^2} \; {\cal{F}}(k) \equiv  A_t\left(\frac{k_*}{a_0}\right)\left(\frac{k}{k_*}\right)^{n_t(k_*) +  \frac{\alpha_t}{2}\ln(k/k_*)
+  \frac{\beta_t}{6}\ln^2(k/k_*) + ...},
\eqn
where $k_*$ denotes the pivot mode,  $a_0$ is the current value of the expansion factor, and  %Thus, we find that
\bqn
\lb{eqA.22}
n_s(k_*) &\equiv& \left. 1+ \frac{d\ln\Delta^2_{\cal{R}}(k)}{d\ln k}\right|_{k = k_*}, \quad 
\alpha_s(k_*) \equiv \left. \frac{d^2\ln\Delta^2_{\cal{R}}(k)}{d\ln k ^2}\right|_{k = k_*}, \quad 
\beta_s(k_*) \equiv \left. \frac{d^3\ln\Delta^2_{\cal{R}}(k)}{d\ln k ^3}\right|_{k = k_*},
\nb\\
n_t(k_*) &\equiv& \left. \frac{d\ln\Delta^2_t(k)}{d\ln k}\right|_{k = k_*}, \quad 
\alpha_t(k_*) \equiv \left. \frac{d^2\ln\Delta^2_{t}(k)}{d\ln k ^2}\right|_{k = k_*}, \quad 
\beta_t(k_*) \equiv \left. \frac{d^3\ln\Delta^2_{t}(k)}{d\ln k ^3}\right|_{k = k_*}.
\eqn
Note that 
\bqn
\lb{eqA.23}
\eta_H = \frac{\ddot{H}}{2H\dot{H}} = \frac{1}{2H}\frac{d}{dt}\left(\ln|\dot H|\right) = \frac{1}{2H}\frac{d}{dt}\ln \left(H^2|\epsilon_H|\right)
= - \epsilon_H + \frac{1}{2}\frac{d}{dN}\left(\ln |\epsilon_H|\right),
\eqn
where $dN \equiv H dt$. Then,  we find that 
\bqn
\lb{eqA.24}
\frac{d\ln|\epsilon_H|}{dN} = 2\left(\epsilon_H + \eta_H\right).
\eqn
On the other hand, at the horizon crossing we have $k = aH$, that is, $\ln k = N + \ln H$, from which we obtain
\bqn
\lb{eqA.25}
\left. \frac{dN}{d\ln k}\right|_{k=k_*} = \left. \left(\frac{d\ln k}{dN}\right)^{-1}\right|_{k=k_*} = \frac{1}{1-\epsilon_H^*},  
\eqn
where $\epsilon_H^* \equiv \epsilon_H(k_*)$. Hence, we find
\bqn
\lb{eqA.26}
\left.\frac{d\ln\Delta^2_{\cal{R}}(k)}{d\ln k}\right|_{k=k_*} &=& \left. \frac{d}{d\ln k}\Big[2\ln H - \ln |\epsilon_H| + \ln{\cal{F}}(k)\Big]\right|_{k=k_*} = \left. \xi_{\text{nBD}}(k_*) + \frac{dN}{d\ln k}\left(-2\epsilon_H - \frac{d\ln|\epsilon_H|}{dN}\right)\right|_{k=k_*} \nb\\
&=& \xi_{\text{nBD}}(k_*) - \frac{2(2\epsilon_H^* + \eta_H^*)}{1-\epsilon_H^*}, 
\eqn
where $\xi_{\text{nBD}}(k_*)$ denotes the non-BD vacuum effects, defined as 
\bqn
\lb{eqA.27}
 \xi_{\text{nBD}}(k_*)  \equiv  \left. \frac{d\ln{\cal{F}}(k)}{d\ln k}\right|_{k = k_*}.
\eqn
If the BD vacuum is chosen, we have $\alpha_k = 1$ and $\beta_k = 1$, so that 
${\cal{F}}(k) = 1$ and $\xi_{\text{nBD}}(k_*) = 0$. 
Inserting the above expressions into Eq.(\ref{eqA.22}), we obtain
\bqn
\lb{eqA.27}
 && n_s(k_*) - 1  =  \xi_{\text{nBD}}(k_*)  - \frac{2(2\epsilon_H^* + \eta_H^*)}{1-\epsilon_H^*}, \\
 &&  n_t(k_*)  =   \xi_{\text{nBD}}(k_*) -  \frac{2\epsilon_H^*}{1-\epsilon_H^*}.
\eqn
It is important to note that {\em all the equations presented so far in this subsection hold for any inflation as long as its background is quasi-de Sitter}.

\subsection{Slow-Roll Inflation}

The slow-roll inflation requires 
\bq
\lb{eqA.31}
\epsilon_H, \; |\eta_H| \ll 1. 
\eq
Then, from Eq.(\ref{eqA.8}) we can see that the second condition of the above equation is equivalent to require $|\ddot\phi| \ll |H\dot\phi|$. As a result, 
Eq.(\ref{eqA.4}) yields 
\bqn
\lb{eqA.32}
\dot\phi \simeq - \frac{V_{,\phi}}{3H}.  
\eqn
On the other hand, the first condition of    Eq.(\ref{eqA.31}) implies $H^2 \gg 4\pi G \dot\phi^2$, so that
\bqn
\lb{eqA.33}
H^2 = \frac{8\pi G}{3}\left(\frac{1}{2}\dot\phi^2 + V(\phi)\right) \simeq \frac{8\pi G}{3}V(\phi).    
\eqn
Hence, we have
\bqn
\lb{eqA.34}
\epsilon_H = - \frac{\dot H}{H^2} \simeq - 4\pi G \dot\phi^2 \frac{3}{8\pi G V(\phi)} \simeq  \frac{1}{16\pi G}\left(\frac{V_{,\phi}}{V}\right)^2 = \epsilon_V.
\eqn
 On the other hand, from Eq.(\ref{eqA.8}) we find that 
 \bqn
\lb{eqA.35}
\eta_H &=& \frac{1}{H}\frac{d}{dt}\ln|\dot\phi| \simeq \frac{1}{H}\frac{d}{dt}\left[\ln\left(- \frac{V_{,\phi}}{3H}\right)\right] 
\simeq \epsilon_H - \frac{1}{H} \left(- \frac{V_{,\phi}}{3H}\right)\frac{V_{,\phi\phi}}{V_{,\phi}} \nb\\
&\simeq&   \epsilon_H  - \frac{1}{8\pi G}\left(\frac{V_{,\phi\phi}}{V}\right) = \epsilon_V - \eta_V.
\eqn
Therefore, with the slow-roll conditions (\ref{eqA.31}), the two sets of parameters are related via the relations
\bqn
\lb{eqA.36}
\epsilon_H \simeq \epsilon_V, \quad  \eta_H \simeq   \epsilon_V - \eta_V.
\eqn
Then, Eq.(\ref{eqA.27}) becomes
\bqn
\lb{eqA.37}
 && n_s(k_*) - 1  \simeq   \xi_{\text{nBD}}(k_*)  - 2\left(3\epsilon_V^* - \eta_V^*\right), \\
 \lb{eqA.38}
 &&  n_t(k_*)  \simeq   \xi_{\text{nBD}}(k_*)  -  2\epsilon_V^*.
\eqn

Planck 2018 chose $k_*/a_0 = 0.05/$Mpc and  obtained the following constraints for the scalar modes \cite{Planck:2018jri}
\bqn
\lb{eqA.22b}
n_s = 0.9587 \pm 0.0056, \quad 
\alpha_s = 0.013 \pm 0.012, \quad 
\beta_s = 0.022 \pm 0.012,
\eqn
from the TT+lowE+lensing data all at $68\%$ CL. 

\subsection{Calculation of the Pivot Mode $k_*$}

To calculate the pivot mode $k_* = k_*^{\text{ob}} a_0$, where $k_*^{\text{ob}} = 0.05\; \text{Mpc}^{-1}$ for the Planck 2018 observations, 
let us first note that at the horizon crossing we have  the following algebraic equations
\bqn
\lb{eqA.38}
 && H_*^2 = \pi m_{\text{PL}}^2 \frac{\epsilon_V^* A_s(k_*)}{{\cal{F}}(k_*)},\\
 \lb{eqA.39}
 &&  \xi_{\text{nBD}}(k_*)  - 2\left(3\epsilon_V^* - \eta_V^*\right) = n_s(k_*) -1, \\
  \lb{eqA.40}
 && k_* = a_* H_*,\\
 \lb{eqA.41}
&& \dot\phi_* \simeq - \frac{V_{,\phi}(V_0,\phi_*)}{3H_*},\\
\lb{eqA.42}
&& H_*^2 \simeq   \frac{8\pi G}{3}V(V_0,\phi_*),\\
\lb{eqA.44}
&& {k_*} =  {a_0}  k_*^{ob}, 
\eqn
where in LQC we have \cite{Zhu:2017jew}
\bqn
\lb{eqA.43}
 {\cal{F}}(k) = 1 + \delta_{\text{PL}}(k) =  1+ \left[1 + \cos\left(\frac{\pi}{\sqrt{3}}\right)\right]{\text{csch}}^2 \left(\frac{\pi k}{\sqrt{6} k_B}\right),\quad k_B \equiv a_B \sqrt{\frac{8\pi \rho_c}{m_{Pl}^2}},
\eqn
with $\text{csch}\; x \equiv (\sinh x)^{-1}$. In addition, because of the  rescaling symmetry, $a(t) \rightarrow a(t)/L$, of the dynamical equations (\ref{friedmann}) and (\ref{klein-gordon}), where $L$ is a real constant, without loss of the generality, we can always set
\bqn
\lb{eqA.45dd}
a_B\left(H_*, \phi_*, \dot{\phi}_*\right) = 1.
\eqn
Note that, once $\left(H_*, \phi_*, \dot{\phi}_*\right)$ are given, the dynamical   equations (\ref{friedmann}) and (\ref{klein-gordon}) will uniquely determine $\left(a_B, \phi_B, \dot{\phi}_B\right)$. So, Eq.(\ref{eqA.45dd}) represents a non-local constraint of the three parameters $\left(H_*, \phi_*, \dot{\phi}_*\right)$ through the field equations.

On the other hand, once $a_B$ is fixed, we can calculate $a_0$ via the relation 
\bqn
\lb{eqA.45gg}
a_0 = a_B e^{N_T} = a_*e^{\hat{N}_T},
\eqn
where
\bqn
\lb{eqA.45}
N_T &\equiv& \ln\left(\frac{a_0}{a_B}\right) = \ln\left(\frac{a_*}{a_B}\cdot \frac{a_0}{a_*}\right) = \ln\left(\frac{a_*}{a_B}\right) + \hat{N}_T, \nb\\
\hat{N}_T &\equiv& \ln\left(\frac{a_0}{a_*}\right) = \ln\left(\frac{a_{\text{end}}}{a_*} \cdot \frac{a_{\text{re}}}{a_{\text{end}}} \cdot \frac{a_0}{a_{\text{re}}}\right) =   N_* + N_{\text{re}} +  \ln\left(\frac{a_0}{a_{\text{re}}}\right).
\eqn
Here $N_*, N_{\text{re}}$ and ${a_0}/a_{\text{re}}$, given respectively by Eqs.(\ref{Tre}), (\ref{Ninf_general}) and (\ref{aoare}), are functions of $(\phi_*, \phi_{\text{end}}, H_*)$. However, from the slow-roll conditions and the definition of the end of inflation, we have
\bqn
\lb{eqA.45aa}
\epsilon_V(\phi_{\text{end}}) = 1,  
\eqn
which uniquely determine $\phi_{\text{end}}$. For example, for the potential given by Eq.(\ref{potential}), we have
\bqn
\lb{eqA.45bb}
4\pi\phi_{\text{end}}^2\left(1+\phi_{\text{end}}^2\right)^2   = 1,  
\eqn
as can be seen from Eq.(\ref{eq3.7}). Then, from Eqs.(\ref{eqA.45gg}) - (\ref{eqA.45aa})  we find that $N_T = N_T(a_*, H_*, \phi_*)$ for any given $(w_{\text{re}}, g_{\text{re}}, T_0, k_*^{\text{ob}})$. Hence, we have
\bqn
\lb{eqA.45hh}
a_0 = a_B e^{N_T} = a_*e^{\hat{N}_T} = a_0\left(a_*, \phi_*, H_*\right).
\eqn
 In summary, for any given $(w_{\text{re}}, g_{\text{re}}, T_0, k_*^{\text{ob}})$,  
we have seven equations, Eqs.(\ref{eqA.38})-(\ref{eqA.44}) and Eq.(\ref{eqA.45dd}),  for the six unknown
\bqn
\lb{eqA.45cc}
\left(a_*, H_*, \phi_*, \dot\phi_*, k_*, V_0\right),
\eqn
once $(A_s, n_s)$ are provided by observations. Thus, the system, consisting of the seven equations, seems overdetermined. However, we find that (\ref{eqA.44}) is not independent and is satisfied once Eqs.(\ref{eqA.38})-(\ref{eqA.42}) and (\ref{eqA.45aa}) hold. Therefore, the  six equations, (\ref{eqA.38}) - (\ref{eqA.42}) and (\ref{eqA.45dd}), uniquely determine the six unknown
(\ref{eqA.45cc}), and the problem becomes deterministic \footnote{In all the  analysis given in this appendix, we assume that the potential $V(\phi)$ depends only on one free parameter, $V_0$. This is the case for various viable potentials \cite{Planck:2018jri}. However, there are cases in which $V(\phi)$ depends on several parameters. In these cases, to determine the free parameters of the potentials, additional physics needs to be taken into account \cite{Kallosh:2025ijd}.}.

\begin{figure}
{\includegraphics[width=0.49\textwidth]{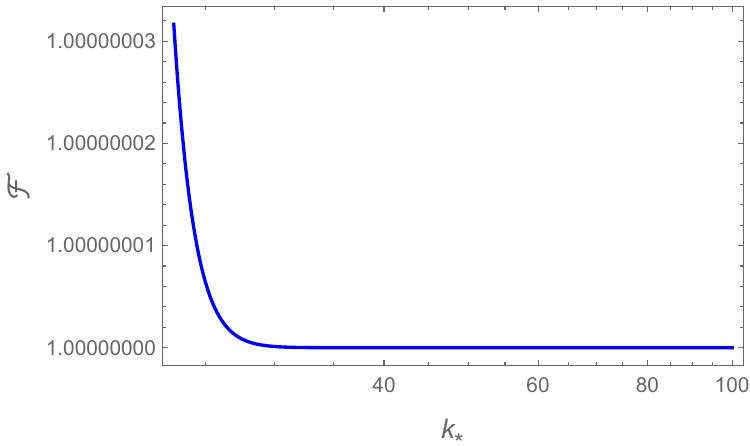}}
{\includegraphics[width=0.49\textwidth]{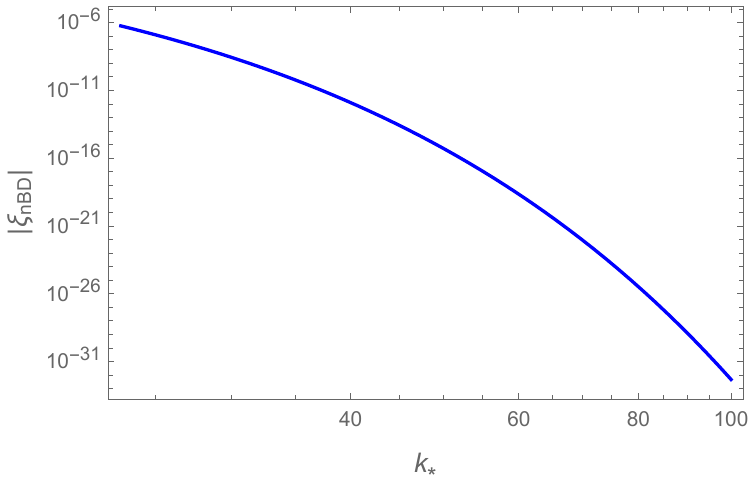}}
\caption{The functions ${\cal{F}}(k)$ and $\xi_{\text{nBD}}(k)$ defined respectively by Eqs.(\ref{eqA.43}) and (\ref{eqA.27}). }
\label{fig8}
\end{figure}

The above analysis provides a routine for computing the values of the relevant physical quantities of the pivot mode $k_*$ at the horizon crossing $t=t_*$, given the form of the inflationary potential. 

\subsection{Quantum Geometric Effects}

To understand the quantum geometric effects, let us first consider the functions ${\cal{F}}(k)$ and $\xi_{\text{nBD}}(k)$ defined respectively by Eqs.(\ref{eqA.43}) and (\ref{eqA.27}), which are plotted in Fig. \ref{fig8}. From this figure it can be seen that the quantum effects are very small (except for the case $k \ll k_B$). Therefore, to the first-order approximation, we can set  ${\cal{F}}(k) = 1$ and $\xi_{\text{nBD}}(k) = 0$. Then, we find
\bqn
\lb{eqA.38b}
 && H_*^2 = \pi m_{\text{PL}}^2 \epsilon_V^* A_s(k_*),\\
 \lb{eqA.39b}
 &&  2\left(\eta_V^* - 3\epsilon_V^*\right) = n_s(k_*) -1, \\
  \lb{eqA.40b}
 && k_* = a_* H_*,\\
 \lb{eqA.41b}
&& \dot\phi_* \simeq - \frac{V_{,\phi}(V_0,\phi_*)}{3H_*},\\
\lb{eqA.42b}
&& H_*^2 \simeq   \frac{8\pi G}{3}V(V_0,\phi_*),\\
 \lb{eqA.44b}
 && a_B\left(a_*, \phi_*, \dot{\phi}_*\right)  =   1.
\eqn

To solve the above equations, our strategy is as follows:

\begin{itemize}
    \item Since $\epsilon_V^*$ and $\eta_V^*$ are functions of $\phi_*$ only, then for any given observational values of $(A_s, n_s)$, from Eq.(\ref{eqA.39b}) we find $\phi_*$. Then, from Eq.(\ref{eqA.38b})
we find $H_{*} = \epsilon_{\pm} \sqrt{\pi m_{\text{PL}}^2 \epsilon_V^* A_s(k_*)}$, where $ \epsilon_{\pm} = \pm 1$.  Then, inserting them into Eq.(\ref{eqA.42b}) we find $V_0$. With such obtained $(H_*, \phi_*, V_0)$, Eq.(\ref{eqA.41b}) will give 
$\dot\phi_*$, which has two solutions, corresponding respectively to the two different choices of $\epsilon_{\pm}$. Therefore, now we find the following four quantities
\bqn
\lb{eqA.60}
\left(H_*, \phi_*, \dot{\phi}_*, V_0\right).
\eqn

\item  With the values of $(\phi_*, \dot\phi_*)$ obtained above, we choose a value of $a_*$, say, $\hat{a}_*$, and then evolute the dynamical equations (\ref{friedmann}) and (\ref{klein-gordon}) back to the bounce time to find 
$(a_B, \phi_B, \dot\phi_B)$. Clearly, such obtained $a_B(\hat{a}_*)$ is usually not equal to 1. We shall keep trying different values of $\hat{a}_*$, until we find that 
$a_B = 1$. This value of $\hat{a}_*$ is the value that we look for, which will be denoted by $a_*$, so we have $a_B\left(a_*, \phi_*, \dot{\phi}_*\right)   = 1$.

It is interesting to note that in the above process, we find that 
such obtained $(\phi_B, \dot\phi_B)$ do not depend on the value of
$\hat{a}_*$. This is expected, as the dynamical equations (\ref{friedmann}) and (\ref{klein-gordon}) have the scaling symmetry, $\hat{a}_* \rightarrow \hat{a}_*/L$, as mentioned above. The above observation also leads to a more efficient way to find ${a}_*$: Choosing a value of  $\hat{a}_*$ together with the values of $(\phi_*, \dot\phi_*, V_0)$ found in Eq.(\ref{eqA.60}) to evolute the dynamical equations (\ref{friedmann}) and (\ref{klein-gordon}) until the bounce to obtain
\bqn
\lb{eqA.62}
\left(\phi_B, \dot{\phi}_B\right).  
\eqn
Then,  taking $a_B = 1$ and $(\phi_B, \dot{\phi}_B)$ given by Eq.(\ref{eqA.62}) as the initial conditions at the bounce, 
evolute the  dynamical equations (\ref{friedmann}) and (\ref{klein-gordon}) from $t_B$ to $t_*$ to find $a_*$. Clearly, such obtained $a_*$ will satisfy Eq.(\ref{eqA.44b}). To make sure that this indeed corresponds to the horizon-cross of the pivot mode $k_*$, we can calculate $(H_*, \phi_*, \dot{\phi}_*)$ and then compare them with these given by Eq.(\ref{eqA.60}). The consistency of these values implies that they are indeed corresponding to the pivot mode. 

\item Once we find the quantities 
\bqn
\lb{eqA.63}
\left(a_*, H_*, \phi_*, \dot{\phi}_*, V_0\right),
\eqn
from Eq.(\ref{eqA.40b}) we can find $k_*$. 

\item  On the other hand, from Eqs.(\ref{eqA.40}) and (\ref{eqA.44})  we obtain  
\bqn
\lb{eqA.63aa}
H_* %= a_0 k_*^{\text{ob}} = e^{N_T} k_*^{\text{ob}} = a_*
= e^{\hat{N}_T(H_*, \phi_*)} k_*^{\text{ob}}, 
\eqn
where $\hat{N}_T$ is given by Eq.(\ref{eqA.45}). Recall that for Planck 2018 the pivot mode $k_*^{\text{ob}} \equiv k_*/a_0$ is chosen as $k_*^{\text{ob}} = 0.05/\text{Mpc}$. Inserting the values 
of $(a_*, k_*, H_*, \phi_*, \dot{\phi}_*, V_0)$  obtained in the above steps into Eq.(\ref{eqA.63aa})  we check the consistency of the system.

\item Now let us consider the quantum effects for which we have ${\cal{F}}(k) \not= 1$. Considering the fact that ${\cal{F}}(k)$ is not significantly different from one, as shown by Fig. \ref{fig8}, we can consider the solutions obtained in the previous steps as the first-order approximation, denoted by $\mathsf{A}_*^{(0)}$,  and then expand them as
\bqn
\lb{eqA.64}
\mathsf{A}_* = \mathsf{A}_*^{(0)} + \delta \mathsf{A}_*,
\eqn
to find $\delta \mathsf{A}_*$ by repeating the above steps, where $\mathsf{A}_* = \left\{{\cal{F}}(k_*), a_*, H_*, \phi_*, \dot{\phi}_*, k_*\right\}$. 

\item Certainly, if we need higher accuracy, we can expand $\mathsf{A}_*$ to high orders, and repeat the above steps to obtain the six physical quantities $\left(a_*, H_*, \phi_*, \dot{\phi}_*, k_*,V_0\right)$ at the horizon crossing. These quantities uniquely determine the whole evolution of the universe, from the contracting phase to the current time through the dynamical equations (\ref{friedmann}) and (\ref{klein-gordon}).

\end{itemize}

It should be noted that when the reheating phase is not taken into account, the e-fold $\hat{N}_T$ becomes a free parameter, and from Eq.(\ref{eqA.63aa}) we can read off $\hat{N}_T$ for the value of $k_*$ found from Eq.(\ref{eqA.63aa}) \cite{Ashtekar:2011rm,Agullo:2013ai}.

%%%%%%%%%%%%%%%%%%%%%%%%%%%%%%%%%

%\newpage
\bibliographystyle{apsrev4-1}
\bibliography{PLP_LQC.bib}

\end{document}